\documentclass[
aps,
prl,
showpacs,
floatfix,
reprint,
twoside,
superscriptaddress
]{revtex4-1}
\usepackage{amsmath}
\usepackage[T1]{fontenc}
\usepackage{fourier}
\usepackage{graphicx}
\usepackage[colorlinks=true,
allcolors=blue,
dvipdfm=true]{hyperref}

\begin{document}

\title{Nematic versus ferromagnetic spin filtering of triplet Cooper pairs in superconducting spintronics}

\author{Andreas Moor}
\affiliation{Theoretische Physik III, Ruhr-Universit\"{a}t Bochum, D-44780 Bochum, Germany}
\author{Anatoly F.~Volkov}
\affiliation{Theoretische Physik III, Ruhr-Universit\"{a}t Bochum, D-44780 Bochum, Germany}
\author{Konstantin B.~Efetov}
\affiliation{Theoretische Physik III, Ruhr-Universit\"{a}t Bochum, D-44780 Bochum, Germany}
\affiliation{National University of Science and Technology ``MISiS'', Moscow, 119049, Russia}

\begin{abstract}
We consider two types of magnetic Josephson junctions~(JJ). They are formed by two singlet superconductors~S and magnetic layers between them so that the JJ is a heterostructure of the S$_{\text{m}}$/n/S$_{\text{m}}$ type, where~S$_{\text{m}}$ includes two magnetic layers with non-collinear magnetization vectors. One layer is represented by a weak ferromagnet and another one---the spin filter---is either conducting strong ferromagnet (nematic or N\nobreakdash-type JJ) or magnetic tunnel barrier with spin-dependent transparency (magnetic or M\nobreakdash-type JJ). Due to spin filtering only fully polarized triplet component penetrates the normal n~wire and provides the Josephson coupling between the superconductors~S. Although both filters let to pass triplet Cooper pairs with total spin~$\mathbf{S}$ parallel to the filter axes, the behavior of nematic and magnetic JJs is completely different. Whereas in the nematic case the charge and spin currents,~$I_{\text{Q}}$ and~$I_{\text{sp}}$, do not depend on mutual orientation of the filter axes, both currents vanish in magnetic~JJ in case of antiparallel filter axes, and change sign under reversing the filter direction. The obtained expressions for~$I_{\text{Q}}$ and~$I_{\text{sp}}$ show clearly a duality between the superconducting phase~$\varphi$ and the angle~$\alpha$ between the exchange fields in the weak magnetic layers.
\end{abstract}

\date{\today}

\maketitle

Triplet Cooper pairing is known to exist in superfluid $^{3}$He~\cite{vollhardt,volovik}. As concerns superconductors, the situation is less clear. Although some indication for the triplet superconductors has been found in a number of completely different classes of materials~\cite{jerome,SrRuO,Mackenzie_RevModPhys.75.657}, a general consensus about the existence of the triplet superconductivity in the organic metals, heavy fermions and other interesting materials investigated from this point of view has not been achieved~\cite{lebed,Mineev}. In principle, the fact that the spins of the fermions of the Cooper pairs are equal to each other does not contradict the Pauli principle because the condensate wave function~$f(\mathbf{p})$ and the order parameter~$\Delta_{\text{tr}}(\mathbf{p})$ in these triplet superconductors are odd functions of the momentum~$\mathbf{p}$. In contrast to the conventional BCS superconductivity, the triplet superconductivity with such a symmetry of the order parameter is sensitive to impurity scattering~\cite{Larkin_Ovchinnikov_1965} and, therefore, it is usually strongly suppressed by disorder.

However, the triplet Cooper pairs can appear already in conventional singlet superconductors provided an external magnetic~($\mathbf{H}$) or an internal exchange~($\mathbf{h}$) field acts on the spins of electrons.\cite{Larkin_Ovchinnikov_1965,Fulde_Ferrell_1964,Fulde_1973,Meservey_1994} A triplet component inevitably arises also in magnetic superconductors~\cite{Bulaevskii_et_al_Adv_Phys}.

The triplet condensate function arising from the singlet superconductivity in the presence of the magnetic or exchange field acting on spins is odd in frequency and therefore may still have an $s$\nobreakdash-wave space symmetry without violating the Pauli principle. It has a component with the $0$\nobreakdash-spin projection ${f_{\text{tr},0} \propto \langle \hat{c}_{\uparrow} \hat{c}_{\downarrow}(t) + \hat{c}_{\downarrow} \hat{c}_{\uparrow}(t) \rangle}$ on the direction of the field~$\mathbf{h}$ or~$\mathbf{H}$ but also the components with spin projection~$\pm 1$. The zero projection component~$f_{\text{tr},0}$ of the condensate function is as sensitive to the exchange or magnetic field as the singlet condensate function ${f_{\text{s}} \propto \langle \hat{c}_{\uparrow} \hat{c}_{\downarrow}(t) - \hat{c}_{\downarrow} \hat{c}_{\uparrow}(t) \rangle}$. It is this type of the triplet pairing that arises in the case of uniform magnetic~$\mathbf{H}$ or exchange~$\mathbf{h}$ field, and has been considered in many works (see, e.g.,~\cite{Larkin_Ovchinnikov_1965,Fulde_Ferrell_1964,Fulde_1973,Bulaevskii_et_al_Adv_Phys}). When the zero projection triplet condensate is created at a superconductor-ferromagnet~(SF) interface, it penetrates the latter over a rather short scale ${\xi_{h} \propto 1 / \sqrt{h}}$ (diffusive case), see reviews Refs.~\onlinecite{BVErmp,Eschrig_Ph_Today,Buzdin_RMP_2005}.

In contrast to the zero projection condensate, the odd frequency triplet pairing with $\pm 1$~components that are proportional to ${f_{\text{tr},+1} \propto \langle c_{\uparrow} c_{\uparrow}(t) \rangle}$ or ${f_{\text{tr},-1} \propto \langle c_{\downarrow} c_{\downarrow}(t) \rangle}$ is rather insensitive to the exchange field and can penetrate a ferromagnet over long distances like the conventional singlet condensate penetrates the normal metal. As has been shown theoretically in Ref.~\cite{Bergeret_Volkov_Efetov_2001} this triplet component arises in the case of a nonuniform~$\mathbf{h}$ and penetrates the ferromagnet~F over a rather long distance. [A little later, a similar idea about the LRTC was considered qualitatively in Ref.~\cite{Kad01}. Contrary to the diffusive case analyzed in Ref.~\cite{Bergeret_Volkov_Efetov_2001} (the mean free path~$l$ is shorter than~$\xi_{h}$) the authors of Ref.~\cite{Kad01} assumed that the length of the magnetic inhomogeneity is shorter than~$l$. They attempted to estimate the amplitude of the LRTC. A microscopic theory of the LRTC for this quasiballistic case has been presented in Ref.~\cite{VE08}.] Since the discovery of the long range triplet odd frequency components~$f_{\text{tr},\pm 1}$, the idea of the penetration of superconductivity through strong ferromagnets has been discussed in numerous publications (see Refs.~\onlinecite{Asano_et_al_2007,Buzdin07,Zaikin08,Linder09,Radovic10,Valls08}, the reviews Refs.~\onlinecite{BVErmp,Eschrig_Ph_Today,Buzdin_RMP_2005} as well as the recent works Refs.~\onlinecite{Halterman_et_al_arXiv_2015,Alidoust_et_al_2015,Leksin_et_al_arXiv_2015,Fominov_et_al_2015,Banerjee_et_al_2014} and references therein). The theoretical research has been followed by experimental confirmation of the existence of the long range triplet superconductivity~\cite{Keizer06,Petrashov06,Birge10,*Birge12,Aarts10,*Aarts11,*Aarts12,Blamire10,*Blamire12,Zabel10,Petrashov11,Kalcheim_et_al_2011,Leksin_et_al_2012,Khaydukov_et_al_2014}.

Note that the triplet component that arises in magnetic superconductors with a spiral~$\mathbf{h}(\mathbf{r})$~\cite{Bulaevskii_et_al_Adv_Phys,Bulaevskii_Rusinov_Kulic_1980} has zero projection of the total spin of triplet Cooper pairs~$\mathbf{S}$ on~$\mathbf{h}(\mathbf{r})$. Only near the surface of the sample, the triplet component with a nonzero projection of~$\mathbf{S}$ on~$\mathbf{h}(\mathbf{r})$ appears; it decays exponentially away from the surface~\cite{Volkov_2008}.

Due to an ability of changing the exchange field direction of the ferromagnets one can filter the spin polarization of the triplet Cooper pairs~\cite{Birge10,*Birge12,Blamire10,*Blamire12} and one may even speak of ``Superconducting spintronics''~\cite{Linder_Robinson_2015}. The possibility of real applications is enforced by the fact that the wave function of odd triplet superconductivity has the $s$\nobreakdash-wave symmetry and, therefore, this triplet superconductivity is not sensitive to non-magnetic disorder. In this situation, it is very important to understand what kind of magnetic filters can be built in order to filter the spin polarization of the currents.

In this Paper, we demonstrate that two types of filtering are possible: ferromagnetic and nematic. A ferromagnetic filter enables one to achieve the full spin polarization, such that the current has only one allowed direction of spin. In contrast, the nematic filter allows to pass the current with spin polarizations both parallel and antiparallel to a certain axis.

Naively, the reason why the triplet pairs in the presence of the exchange field are not destroyed by the Zeeman interaction, is the fact that the total spin of the pair is oriented along the field and is not affected by the latter. In principle, this is really the case for a very strong exchange field~$\mathbf{h}$ when the exchange energy~$\mu_{\text{B}} h$ is comparable to the Fermi energy~$\varepsilon_{\text{F}}$. However, in moderate exchange fields~$h$, such that~$\mu_{\text{B}} h$ is considerably smaller than~$\varepsilon_{\text{F}}$, the superconducting pairing is weakly sensitive to the exchange field not only when the total spin of the Cooper pair is parallel but also when it is antiparallel to the latter~\cite{Bergeret_Volkov_Efetov_2001,BVErmp}. It is natural to classify the corresponding filterings as \emph{ferromagnetic}~(M) and \emph{nematic}~(N).

We investigate the difference between the two types of the filtering considering M\nobreakdash- and N\nobreakdash-Josephson junctions~(JJ). We show that, although many effects are similar in these junctions, drastically different phenomena are also possible. For example, \emph{fully polarized} triplet Cooper pairs appear in the normal metal~(n) inside the M\nobreakdash-type of JJs (ferromagnetic order), while there are triplet pairs with spin up \emph{and} spin down in the normal metal in the N\nobreakdash-type of JJs (nematic order).

\begin{figure}
  \centering
  \includegraphics[width=1.0\columnwidth]{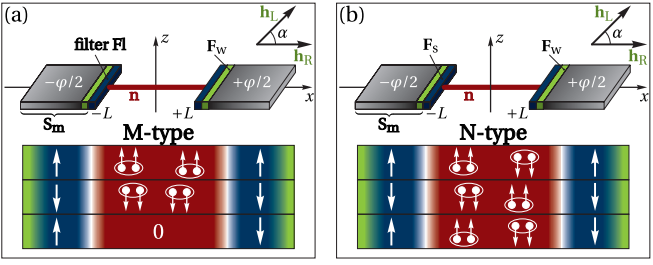}
  \caption{Schematic representation of the system under consideration and classification of the triplet component relative to the orientation of the filters in corresponding JJs, i.e., case~(a) for ``ferromagnetic''~(M) type, and case~(b) for ``nematic''~(N) type as follows from the expressions for the Josephson charge and spin currents, Eqs.~\ref{8}--\ref{11} (see text).} \label{fig:setup}
\end{figure}

We consider two types of symmetric ``magnetic'' Josephson junctions: a)~S$_{\text{m}}$/Fl/n/Fl$^{\prime}$/S$_{\text{m}}^{\prime}$ and b)~S$_{\text{m}}$/F$_{\text{s}}$/n/F$_{\text{s}}^{\prime}$/S$_{\text{m}}^{\prime}$. Both types of JJs are shown in the upper parts of the corresponding Figs.~\ref{fig:setup}~(a) and~\ref{fig:setup}~(b). The M\nobreakdash-type JJ consists of a normal wire or film which connects two~S$_{\text{m}}$ and~S$_{\text{m}}^{\prime}$ reservoirs, representing superconductors with exchange fields~$\mathbf{h}_{\text{R},\text{L}}$ defining the $(x,y)$~plane. The superconductors~S$_{\text{m}}$ may be realized in the form of a superconductor/weak ferromagnet bilayer, S/F$_{\text{w}}$, with an effective exchange field ${\mathbf{h} = \mathbf{h}_{\text{F}} d_{\text{F}} / (d_{\text{F}} + d_{\text{S}})}$, where~$\mathbf{h}_{\text{F}}$ is an exchange field in~F$_{\text{w}}$ and~$d_{\text{F},\text{S}}$ are the thicknesses of the~F$_{\text{w}}$ and S~films, respectively~\cite{Bergeret_Volkov_Efetov_2001_b}. The n~wire is separated from~S$_{\text{m}}$ by a tunnel spin-active barrier~Fl~(filter) with a spin-dependent transparency. Such kinds of magnetic insulators have been already used in experiments, e.g., on ultrathin EuO~films~\cite{Moodera08}. Penetration of Cooper pairs through~Fl is taken into account via boundary conditions. We consider the case when the filter~Fl passes electrons with only one spin direction collinear with the $z$~axis defined by the vector product of the exchange field vectors in the S$_{\text{m}}$~superconductors, i.e., ${\mathbf{z} \propto \mathbf{h}_{\text{R}} \times \mathbf{h}_{\text{L}}}$; in other words, the filter axis is perpendicular to~$\mathbf{h}_{\text{R},\text{L}}$. This means that some triplet Cooper pairs that arise in S$_{\text{m}}$ have a nonzero projection of the total spin~$\mathbf{S}$ onto the $z$~axis and, therefore, can penetrate through the filters.

The N\nobreakdash-type JJ is similar to the M\nobreakdash-type, but the spin-active barriers are replaced by strong diffusive ferromagnetic layers with magnetization~$\mathbf{M}_{\text{s}}$, so that only triplet component with spins collinear with the $z$~axis can penetrate the n~wire.

We employ the well developed method of quasiclassical Green's functions that can be used in problems where quantum effects on the scale of Fermi wave length are unimportant~\cite{RammerSmith,LO,BelzigRev,Kopnin}. In the considered case, when not only singlet, but also short- and long-range triplet components exist in the system, quasiclassical Green's functions~$\hat{g}$ are matrices in the particle-hole and spin spaces with basis represented by tensor products ${\hat{X}_{ik} = \hat{\tau}_{i} \cdot \hat{\sigma}_{k}}$, where the Pauli matrices~$\hat{\tau}_{i}$ and~$\hat{\sigma}_{k}$ (${i,k=1,2,3}$) and the unit matrices~$\hat{\tau}_{0}$ and~$\hat{\sigma}_{0}$ operate in the particle-hole and spin space, respectively (see also Ref.~\cite{BVErmp}).

We study the case of diffusive conductors that corresponds to most experimental structures. This implies that the mean free path~$l$ is much shorter than the coherence length~${\xi_{T} = \sqrt{D/\pi T}}$ with the diffusion coefficient ${D = v_{\text{F}} l/3}$, the Fermi velocity~$v_{\text{F}}$ and the temperature~$T$. In the considered equilibrium case, it is sufficient to find the Green's functions in the Matsubara representation~$\hat{g}(\omega)$. In the n~or F~wire along the $x$\nobreakdash-direction, these matrix functions obey the generalized Usadel equation~\cite{Usadel,BVErmp,Buzdin_RMP_2005}
\begin{equation}
    - \partial_{x}(\hat{g} \partial_{x} \hat{g}) + \kappa_{\omega}^{2} [\hat{X}_{30} \,, \hat{g}] + i \kappa_{H}^{2} [\hat{X}_{33} \,, \hat{g}] = 0 \,,
    \label{1}
\end{equation}
and the normalization condition ${\hat{g} \cdot \hat{g} = 1}$, where ${\kappa_{\omega}^2 = |\omega| / D}$, ${\kappa_{H}^2 = \mathrm{sgn}(\omega) |\mathbf{H}| / D}$, and ${\omega = (2n+1) \pi T}$ is the Matsubara frequency. The coefficient~$\kappa_{H}$ is not zero only in the strong ferromagnet~F$_{\text{s}}$ with an exchange field~$\mathbf{H}$ (the N\nobreakdash-type JJ). In case of the M\nobreakdash-type JJ, the third term in Eq.~(\ref{1}) should be dropped. [Note that the representation of the matrix Green's functions~$\hat{g}$ differs somewhat from those ~$\hat{g}_{\text{BVE}}$ used in Refs.~\cite{Bergeret_Volkov_Efetov_2001,BVErmp}. We employed the transformation suggested by Ivanov and Fominov~\cite{IvanovFomin} so that ${\hat{g} = \hat{U} \cdot \hat{g} \cdot \hat{U}^{\dagger}}$ with ${\hat{U} = (1/2)(1 + i \hat{X}_{33}) \cdot (1 - i \hat{X}_{03})}$.]

The boundary conditions can be written as (see Refs.~ \cite{EschrigBC13,*EschrigBC13a,EschrigBC15} and Eq.~(4.7) in Ref.~\cite{Bergeret12b})
\begin{equation}
    \hat{g} \partial_{x} \hat{g}_{|_{x = \pm L}} = \pm \kappa_{\text{b}} [\hat{g} \,, \hat{\mathrm{\Gamma}} \hat{G} \hat{\mathrm{\Gamma}}]_{|_{x = \pm L}} \,,
    \label{2}
\end{equation}
where ${\kappa_{\text{b}} = (\sigma R_{\text{b}})^{-1}}$ with the conductivity of the n~wire~$\sigma$ and the n\nobreakdash-Fl or n\nobreakdash-F$_{\text{s}}$ interface resistance at~$\pm L$ per unit area~$R_{\text{b}}$ assumed to be equal for left and right banks. The matrix transmission coefficient~$\hat{\mathrm{\Gamma}}_{\nu}$ (with ${\nu = \pm L}$ or R/L) describes the electron transmission with a spin-dependent probability~$\mathcal{T}_{\uparrow, \downarrow}$. If the filters let to pass only electrons with spins aligned parallel to the $z$~axis, then ${\hat{\mathrm{\Gamma}}_{\nu} = \mathcal{T}_{\nu} + \mathcal{U}_{\nu} \hat{X}_{33}}$ so that the probability for an electron with spin up (down) to pass into the n~wire is ${\mathcal{T}_{\uparrow, \downarrow} = \mathcal{T}_{\nu} + \zeta_{\nu} \mathcal{U}_{\nu}}$ with  ${\zeta_{\nu} = +1}$ if the filter passes only spin-up electrons, whereas ${\zeta_{\nu} = -1}$ means that the filter passes spin-down electrons only.

We assume that the coefficients~$\mathcal{T}$ and~$\mathcal{U}$ are normalized, i.e., ${\mathcal{T}^{2} + \mathcal{U}^{2} = 1}$. Therefore, the condition ${\mathcal{T}_{\nu} = \zeta_{\nu} \mathcal{U}_{\nu}}$ means that electrons with only one spin orientation are allowed to pass through the filter. The form of the matrix~$\hat{\mathrm{\Gamma}}$ with arbitrary axis of spin selection can be found by using the rotation transformation ${\hat{\mathrm{\Gamma}}_{\beta} = \hat{R}_{\beta} \cdot \hat{\mathrm{\Gamma}} \cdot \hat{R}_{\beta}^{\dagger}}$ with the rotation matrix ${\hat{R}_{\beta, j} = \cos(\beta/ 2) + i \sin(\beta/ 2) \hat{X}_{0j}}$, where the subindex~$j$ denotes the axis of rotation and~$\beta$ is the angle of rotation. In the N\nobreakdash-type JJ the matrix~$\hat{\mathrm{\Gamma}}_{\nu}$ does not depend on spins, such that ${\hat{\mathrm{\Gamma}}_{\nu} = 1}$.

Quasiclassical Green's function matrices~$\hat{G}_{\nu}$ in superconductors~S$_{\nu}$ in presence of an exchange field~$\mathbf{h}_{\nu}$ oriented along the $z$~axis [${\mathbf{h}_{\nu} = (0, 0 , h_{\nu})}$] have the form (dropping~$\nu$ for brevity)
\begin{equation}
    \hat{G} = g_{+} \hat{X}_{30} + g_{-} \hat{X}_{33} + \hat{F} \,. \label{3}
\end{equation}
They contain the normal part (the first two terms) and the anomalous (Gor'kov's) part,
\begin{equation}
\hat{F} = \hat{X}_{10} f_{+} + \hat{X}_{13} f_{-} \,, \label{F}
\end{equation}
where ${f_{\pm} = (1/2) [f(\omega + i h) \pm f(\omega - i h)]}$ with ${f(\omega) = \Delta / \sqrt{\omega^{2} + \Delta^{2}}}$. The functions~$g_{\pm}$ are obtained from~$f_{\pm}$ using the orthogonality condition.

The first term in the expression for~$\hat{F}$, Eq.~(\ref{F}), represents the singlet component and the second term describes the triplet component with zero projection of the total spin of a Cooper pair on the direction of the exchange field~$\mathbf{h}$. The triplet component~$f_{-}$ turns to zero at ${h = 0}$ and, as follows from the Pauli principle, is an odd function of~$\omega$.

Presence of superconducting phase~$\chi$ is introduced via a gauge transformation ${\hat{F}_{\chi} = \hat{S}_{\chi} \cdot \hat{F} \cdot \hat{S}_{\chi}^{\dagger}}$, with the unitary matrix ${\hat{S}_{\chi} = \cos (\chi /2) + i \hat{X}_{30} \sin (\chi /2)}$. Moreover, with the help of the rotation matrix~$\hat{R}_{\beta, j}$ one can find the matrix~$\hat{F}_{\beta}$ for an arbitrary orientation of the exchange field ${\mathbf{h} = \{ h_{i} \}}$.

In the following, we concentrate on the case when the exchange field~$\mathbf{h}$ is perpendicular to the magnetization~$\mathbf{M}_{\text{s}}$ in the strong ferromagnet, i.e.,~${\mathbf{h} \perp \mathbf{M}_{\text{s}}}$. In the general case, the obtained results depend on the cosine of the angle between these vectors. Thus, the Green's function in the $\nu$\nobreakdash-th superconductor with the phase~${\chi_{\nu} = \pm \varphi/2}$ and with the exchange field lying in the $(x,y)$~plane and setting up the angle~$\alpha$ with the $x$~axis reads ${\hat{G}_{\nu} = \hat{R}_{\alpha, 3} \hat{R}_{-\pi/2, 2} \hat{S}_{\varphi} \hat{G}_0 \hat{S}_{\varphi}^{\dagger} \hat{R}_{-\pi/2, 2}^{\dagger} \hat{R}_{\alpha, 3}^{\dagger}}$.

Using the developed formalism we write the Josephson~$I_{\text{Q}}$ and spin~$I_{\text{sp}}$ currents through the interface~${\nu = \pm L}$ in the form
\begin{align}
    I_{\text{Q},\nu} &= \sigma e^{-1} 2 \pi i T \sum_{\omega} \mathrm{Tr}\{ \hat{X}_{30} \cdot \hat{g} \partial_{x} \hat{g} \}_{|{\nu}}
    \label{4} \\
    I_{\text{sp},\nu} &= \mu_{\text{B}} \sigma e^{-2} 2 \pi i T \sum_{\omega} \mathrm{Tr}\{ \hat{X}_{03} \cdot \hat{g} \partial_{x} \hat{g} \}_{|{\nu}}
    \label{4a}
\end{align}
where $\mu_{\text{B}}$~is the effective Bohr magneton and $e$\nobreakdash---the elementary charge.

In order to calculate the Josephson current, one should determine the condensate function~$\hat{f}$ in the normal wires, which is a part of the full function ${\hat{g} = \hat{g}_{\text{n}} + \hat{f}}$, where $\hat{g}_{\text{n}}$ is the normal Green's function diagonal in the particle-hole space. This can easily be done for the case of small S$_{\text{m}}$/Fl (respectively S$_{\text{m}}$/F$_{\text{s}}$) interface transparencies, i.e., when the interface resistance is much larger than the resistance of the n~film of the length ${\xi_{T} = \sqrt{D / \pi T}}$, which corresponds to the inequality ${R_{\text{b}} \sigma / \xi_{T} \gg 1}$.

The rather technical but straight forward solution of the Usadel equation with boundary conditions in linearized form (see \emph{Supplemental Material}) yields, with account for the formulas for the Josephson~$I_{\text{Q}}$ and the spin~$I_{\text{sp}}$ currents, Eqs.~(\ref{4}) and~(\ref{4a}), following expressions.
\begin{enumerate}
\item[a)] For the M\nobreakdash-type JJ we find
\begin{align}
    I_{\text{Q}} &= - I_{\text{c}} \zeta^2 \sin (\varphi + \zeta \alpha) \,, \label{8} \\
    I_{\text{sp}} &= - \mu_{\text{B}} e^{-1} I_{\text{c}} \zeta^2 \sin (\alpha + \zeta \varphi) \,, \label{9}
\end{align}
The coefficient $\zeta \equiv (\zeta_{\text{R}} + \zeta_{\text{L}})/2$ equals zero if the right and left filters select Cooper pairs with opposite spin orientations and equals~$\pm 1$ if both the filters let to pass Cooper pairs with the same spin directions parallel or antiparallel to the $z$~axis.

\item[b)] For the N\nobreakdash-type JJ we obtain the expressions
\begin{align}
    I_{\text{Q}} &= - I_{\text{c}} \sin (\varphi) \cos(\alpha) \,, \label{10} \\
    I_{\text{sp}} &= - \mu_{\text{B}} e^{-1} I_{\text{c}} \sin (\alpha) \cos(\varphi) \,, \label{11}
\end{align}
that do not contain~$\zeta$. Equations~(\ref{10}) and~(\ref{11}) show that in this case the filters let the spin of the passing current be both---parallel and antiparallel to the direction of~$\mathbf{h}$ with equal probability.
\end{enumerate}

In equations~(\ref{8})\nobreakdash--(\ref{11}) the critical current~$I_{\text{c}}$ equals ${I_{\text{c}} = \lambda I_0}$, where ${\lambda = - \frac{\sum_{\omega} f_{-}^{2}(\omega) \kappa_{\omega}^{-1} \exp(- 2 L \kappa_{\omega}) }{\sum_{\omega} f^{2}(\omega) \kappa_{\omega}^{-1} \exp(- 2 L \kappa_{\omega}) }}$ with~$f_{-}(\omega)$ and~$f(\omega)$ defined in Eq.~(\ref{F}), and~$I_0$ is the critical current in a usual S/n/S Josephson junction.

Equations~(\ref{8})\nobreakdash--(\ref{11}) represent the main results of this Paper. Note that at ${\alpha = 0}$, the Josephson current in both cases has opposite sign as compared to~$I_{\text{Q}}$ in a usual S/n/S JJ, because the amplitude of the triplet component~${f_{-}(\omega) = i \Im f(\omega + i h)}$ is a purely imaginary quantity and, therefore, ${I_{\text{Q}} \sim I_0 \sum_{\omega} f_{-}^{2}}$ is negative. Furthermore, there is an essential difference between the M- and N\nobreakdash-types of the junctions.

Provided the filters (magnetic insulators) let to pass electrons with antiparallel directions of spins (${\zeta = 0}$) both currents vanish in the M\nobreakdash-type JJ. On the contrary, in the N\nobreakdash-type JJ, the currents do not depend at all on the direction of the exchange field~$\mathbf{H}$ in strong ferromagnets, i.e., provided the vectors~$\mathbf{H}$ are parallel to the $z$~axis, the LRTC Cooper pairs with both spin orientations---parallel and antiparallel to the $z$~axis---propagate through the n~wire (nematic filtering). If ${\alpha = 0}$, the Josephson current~$I_{\text{Q}}$ does not depend on the direction of filter axes (${\zeta > 0}$ or ${\zeta < 0}$), whereas in the M\nobreakdash-type JJ, the spin current changes sign. Physically, this means that Cooper pairs with the total spin \emph{either} up \emph{or} down move in the JJ at positive or negative~$\zeta$ and this kind of spin-active interface can be used as a \emph{complete} filter for triplet Cooper pairs. In the N\nobreakdash-type JJ, the spin current vanishes at ${\alpha = 0}$ meaning that, regardless of the direction of~$\mathbf{H}$, Cooper pairs with spin up \emph{and} spin down move in the n~wire with equal probability and therefore a layer of strong ferromagnet \emph{cannot} serve as a filter for only one spin direction of triplet Cooper pairs and we classify this process as \emph{incomplete filtering}. The discussed difference is summarized in the lower parts of corresponding Figs.~\ref{fig:setup}~(a) and~(b) and supporting technical details are provided in the \emph{Supplemental Material}.

One can say that a current carrying M\nobreakdash-type JJ is analogous to the A\nobreakdash-phase of superfluid~$^{3}$He (triplet pairs with a nonzero magnetic moment), whereas a current carrying N\nobreakdash-type JJ resembles the B\nobreakdash-phase of~$^{3}$He (triplet pairs with zero moment)~\cite{Mineev,vollhardt,volovik}.

Equations~(\ref{8})\nobreakdash--(\ref{11}) manifest a duality between the phase of the superconducting OP~$\varphi$ and the angle~$\alpha$ between the magnetization vectors in~S$_{\text{m}}$---the expression for the current~$I_{\text{Q}}$ is converted into the expression for the spin current~$I_{\text{sp}}$ upon replacement ${\varphi \leftrightarrow \alpha}$ and ${e \leftrightarrow \mu_{\text{B}}}$. This duality allows for prediction of Josephson-like effects in purely magnetic systems~\cite{Larkin90,Flavio04,Tremblay10,Moor12,*Moor_Volkov_Efetov_SUST,Nakata_et_al_2014,Chen_Horsch_Manske_2014}. Presence of a non-zero Josephson current in the M\nobreakdash-type JJs at ${\alpha \neq 0}$ and in absence of a phase difference (spontaneous current) means that we deal with the so-called $\varphi$\nobreakdash-contact. These types of JJs may have different mechanisms and have been discussed in several papers~\cite{Buzdin08,Goldobin12,*Goldobin13,Moor13}.

The different dependence of the charge and spin currents,~$I_{\text{Q,sp}}$, on the phase difference~$\varphi$ and the angle~$\alpha$ can be verified experimentally. As follows from Eqs.~(\ref{8})\nobreakdash--(\ref{11}), a spontaneous current may arise at zero phase difference if the angle~$\alpha$ is not zero. This current can be measured as it was done in the experiment by Bauer~\emph{et al.}~\cite{Bauer_et_al_2004} on an S/F/S junction. In this experiment, superconductors~S were connected by a superconducting loop. If the S/F/S junction was in the $\pi$\nobreakdash-state (or in the $\phi$\nobreakdash-state with ${\phi \neq 0}$), a circulating current arose in the loop. By measuring the circulating current one can establish the dependence of the currents~$I_{\text{Q,sp}}$ on~$\varphi$ and~$\alpha$. The phase difference~$\varphi$ can be varied by an external magnetic field~$\mathbf{H}_{\text{ext}}$: ${\varphi/2 \pi = \Phi_{\text{ext}}/ \Phi + n}$, where~${\Phi_{\text{ext}} = |\mathbf{H}_{\text{ext}}| W}$ is the magnetic flux in the loop, $W$---the square of the loop and~$n$---an integer. Note that the direct measuring of the spin current is a more difficult task because this current is not conserved---it decays to zero in the singlet superconductors.

Note that the formula for Josephson current~$I_{\text{Q}}$, Eq.~(\ref{8}), looks similar to the second term in the corresponding equation [Eq.~(8)] in the paper by Grein~\emph{et al.}~\cite{Grein_et_al_2009}, where the propagation of the LRTC in a strong ferromagnet with a large Zeeman splitting of energy levels has been considered (see also Ref.~\onlinecite{Braude07}).

In conclusion, we have considered the dc Josephson effect in two different S$_{\text{m}}$/n/S$_{\text{m}}$ Josephson junctions with a coupling due to penetration of triplet Cooper pairs generated by two ``magnetic'' superconductors~S$_{\text{m}}$ from one superconductor to the other. Calculating Josephson and spin currents we have found that the results are in disaccord with the ``paradigm'' of using a metallic ferromagnetic layer as the filter for selecting one particular direction of spin projection of triplet Cooper pairs. Actually, there is an essential difference in using this kind of filter as compared to a spin-active interface filter, e.g., a magnetic insulator with a strongly spin dependent transparency. In the former case, one obtains the triplet Cooper pairs with the both directions of the total spin, which enables one selecting a certain orientation of the total spin (``easy axis'') but not the direction---a nematic phase, whereas in the latter case, one can really select a preferred direction of the total spin (magnetic moment of the Cooper pair), thus realizing a ferromagnetic phase of the junction. Experimentally, these situations can be distinguished by performing measurements of Josephson current.

In the nematic case, the Josephson current does not depend on mutual orientations of magnetic moments in the strong ferromagnets, while in the ferromagnetic case, it turns to zero if orientations of the exchange fields in magnetic insulating filters are antiparallel to each other. The considered effects in nematic-type of Josephson contacts can be observed on setups used in the current experiments like those presented in Refs.~\cite{Birge10,*Birge12,Blamire10,*Blamire12,Zabel10}. As regards ferromagnetic-type Josephson contacts, one needs to modify the existing multilayered Josephson junctions by replacing layers of strong ferromagnets with ferromagnetic tunnel barriers. Such materials as DyN, GdN or CoFe$_2$O$_4$, NiFe$_2$O$_4$, BiMnO$_3$, NiMn$_2$O$_4$, CoCr$_2$O$_4$ and Sm$_{0.75}$Sr$_{0.25}$MnO$_3$~\cite{Muduli_et_al_arXiv_2015} may serve for this purpose. Spin polarization in some these materials can reach 90--100\%.

Moreover, accompanying spin currents can also be controlled using the discovered duality between the latter and the Josephson currents, thus leading to an intelligent implementation of related effects in promising spintronic devices based on superconductors~\cite{Linder_Robinson_2015,Eschrig_Review_arXiv_2015}.

\acknowledgments

\paragraph{Acknowledgments.} We appreciate the financial support from the DFG via the Projekt~EF~11/8\nobreakdash-2; K.~B.~E.~gratefully acknowledges the financial support of the Ministry of Education and Science of the Russian Federation in the framework of Increase Competitiveness Program of  NUST~``MISiS'' (Nr.~K2-2014-015).

\section{Supplemental Material}

\paragraph{Solution of the Usadel equation.} In order to calculate the Josephson current, one should determine the condensate function~$\hat{f}$ in the normal wires, which is a part of the full function ${\hat{g} = \hat{g}_{\text{n}} + \hat{f}}$, where $\hat{g}_{\text{n}}$ is the normal Green's function diagonal in the particle-hole space. This can easily be done for the case of small S$_{\text{m}}$/Fl (respectively S$_{\text{m}}$/F$_{\text{s}}$) interface transparencies, i.e., when the interface resistance is much larger than the resistance of the n~film of the length ${\xi_{T} = \sqrt{D / \pi T}}$, which corresponds to the inequality ${R_{\text{b}} \sigma / \xi_{T} \gg 1}$.

Considering this limit, we linearize Eqs.~(1) and~(2) with respect to small condensate function~$\hat{f}$ taking into account that, in the zeroth approximation, ${\hat{g}_{0} = \mathrm{sgn}(\omega) \hat{\mathrm{X}}_{30}}$. The linearized Usadel equation and the boundary conditions acquire the form
\begin{equation}
    -\partial_{xx}^{2} \hat{f} + \kappa_{\omega}^{2} \hat{f} + i \kappa_{H}^{2} ( \hat{X}_{03} \cdot \hat{f} - \hat{X}_{30} \cdot \hat{f} \cdot \hat{X}_{33} ) = 0 \,,
    \label{5}
\end{equation}
\begin{equation}
    \partial_{x} \hat{f} = \pm \kappa_{\text{b}} \big( \hat{\mathrm{\Gamma}} \hat{F}_{\chi} \hat{\mathrm{\Gamma}} \big)_{|_{x = \pm L}} \,,
    \label{6}
\end{equation}
where ${\hat{F}_{\chi}(\pm L) = \hat{S}_{\chi} (\pm L) \cdot [\hat{X}_{12} \cos (\alpha / 2) \pm \hat{X}_{11} \sin (\alpha / 2)]}$ and~$\alpha $ is the angle between the vectors of the exchange fields~$\mathbf{h}_{\text{R}}$ and~$\mathbf{h}_{\text{L}}$ (see Fig.~1 of the main text). Equation~(\ref{5}) can be solved easily in a general case but for brevity we present here a solution in the n~wire for the case ${L \gg \xi_{T}}$ when this solution is a sum of two exponentially decaying functions~\cite{Likharev}
\begin{equation}
\hat{f}_{\text{LRTC}} = \hat{f}_{\text{R}} \exp [- \sqrt{2} \kappa_{\omega} (L - x) ] + \hat{f}_{\text{L}} \exp [ - \sqrt{2} \kappa_{\omega} (L + x) ] \,,
\label{7}
\end{equation}
with amplitudes~$\hat{f}_{\text{L},\text{R}}$ having different shapes depending on the JJ~type.

Considering the M\nobreakdash-type JJ, we are interested only in the case ${\mathcal{T}_{\nu} = \pm \mathcal{U}_{\nu}}$ when the singlet component does not penetrate through the filters. Indeed, calculating the product ${\hat{\mathrm{\Gamma}}_{\nu} \cdot \hat{F}_{\nu, \chi} \cdot \hat{\mathrm{\Gamma}}_{\nu} = \hat{S}_{\chi} \cdot \hat{\mathrm{\Gamma}}_{\nu} \cdot \hat{F}_{\nu, \chi} \cdot \hat{\mathrm{\Gamma}}_{\nu} \cdot \hat{S}_{\chi}^{\dagger}}$ we find for the singlet component ${\hat{\mathrm{\Gamma}}_{\nu} \cdot \hat{X}_{10} f_{+} \cdot \hat{\mathrm{\Gamma}}_{\nu} \propto (\mathcal{T}_{\nu}^{2} - \mathcal{U}_{\nu}^{2}) = }0$ if ${\mathcal{T}_{\nu} = \pm \mathcal{U}_{\nu}}$, while for the triplet component we have ${\hat{\mathrm{\Gamma}}_{\nu} \cdot \hat{X}_{13} f_{-} \cdot \hat{\mathrm{\Gamma }}_{\nu} \neq 0}$. Here, we obtain
\begin{equation}
    \hat{f}_{\text{R}/\text{L}} = \kappa_{\omega} \kappa_{\text{b}}^{-1} \hat{S}_{\varphi}(\pm L) \big[ \hat{X}_{|| \text{R}/\text{L}} \cos (\alpha / 2) \pm \sin (\alpha / 2) \hat{X}_{\perp \text{R}/\text{L}} \big] \,, \label{f_M_case}
\end{equation}
where we defined the matrices~${\hat{X}_{|| \text{R}/\text{L}} = \hat{X}_{11} - \zeta_{\text{R}/\text{L}} \hat{X}_{22}}$ and~${\hat{X}_{\perp \text{R}/\text{L}} = \hat{X}_{12} - \zeta_{\text{R}/\text{L}} \hat{X}_{21}}$. As is easily seen, the matrices~$\hat{X}_{||}$ and~$\hat{X}_{\perp}$ correspond, for ${\zeta_{\text{R}/\text{L}} = + 1}$, to propagators of the form~$\langle \hat{c}_{\uparrow} \hat{c}_{\uparrow} (t) \rangle$ and~$\langle \hat{c}_{\downarrow} \hat{c}_{\downarrow} (t) \rangle$, respectively, i.e., they describe the triplet pairs with spin up and spin down, and vice versa for ${\zeta_{\text{R}/\text{L}} = - 1}$.

In the case of the N\nobreakdash-type JJ, we solve Eq.~(\ref{5}) in the~F$_{\text{s}}$ film using the boundary conditions Eq.~(\ref{6}) with ${\hat{\mathrm{\Gamma}} = 1}$ matching the solution given by Eq.~(\ref{7}) in the n~wire. Note that there is no difference between the n and the F~wire for the LRTC~component because it contains the matrices~$\hat{X}_{12,21}$ or~$\hat{X}_{11,22}$ which commute with the matrix~$\hat{X}_{33}$ in Eq.~(\ref{5}) turning the third term to zero. The singlet ($\hat{X}_{10}$) and short range triplet ($\hat{X}_{13}$) components decay fast in the F$_{\text{s}}$~film over the length~${\xi_{H} = \kappa_{H}^{-1}}$ and do not penetrate into the n~wire. For amplitudes of the LRTC we obtain
\begin{equation}
    \hat{f}_{\text{R}/\text{L}} = \kappa_{\omega} \kappa_{\text{b}}^{-1} \hat{S}_{\varphi}(\pm L) \big[ \hat{X}_{11 \text{R}/\text{L}} \cos (\alpha / 2) \pm \sin (\alpha / 2) \hat{X}_{12 \text{R}/\text{L}} \big] \,, \label{f_N_case}
\end{equation}
and it is easily seen that, in contrast to previous case, in this junction we obtain a mixture of triplet Cooper pairs with both orientations of the total spin, since ${\hat{X}_{11,12} \propto \langle \hat{c}_{\uparrow} \hat{c}_{\uparrow} (t) \rangle + \langle \hat{c}_{\downarrow} \hat{c}_{\downarrow} (t) \rangle}$.


\begin{thebibliography}{75}%
\makeatletter
\providecommand \@ifxundefined [1]{%
 \@ifx{#1\undefined}
}%
\providecommand \@ifnum [1]{%
 \ifnum #1\expandafter \@firstoftwo
 \else \expandafter \@secondoftwo
 \fi
}%
\providecommand \@ifx [1]{%
 \ifx #1\expandafter \@firstoftwo
 \else \expandafter \@secondoftwo
 \fi
}%
\providecommand \natexlab [1]{#1}%
\providecommand \enquote  [1]{``#1''}%
\providecommand \bibnamefont  [1]{#1}%
\providecommand \bibfnamefont [1]{#1}%
\providecommand \citenamefont [1]{#1}%
\providecommand \href@noop [0]{\@secondoftwo}%
\providecommand \href [0]{\begingroup \@sanitize@url \@href}%
\providecommand \@href[1]{\@@startlink{#1}\@@href}%
\providecommand \@@href[1]{\endgroup#1\@@endlink}%
\providecommand \@sanitize@url [0]{\catcode `\\12\catcode `\$12\catcode
  `\&12\catcode `\#12\catcode `\^12\catcode `\_12\catcode `\%12\relax}%
\providecommand \@@startlink[1]{}%
\providecommand \@@endlink[0]{}%
\providecommand \url  [0]{\begingroup\@sanitize@url \@url }%
\providecommand \@url [1]{\endgroup\@href {#1}{\urlprefix }}%
\providecommand \urlprefix  [0]{URL }%
\providecommand \Eprint [0]{\href }%
\providecommand \doibase [0]{http://dx.doi.org/}%
\providecommand \selectlanguage [0]{\@gobble}%
\providecommand \bibinfo  [0]{\@secondoftwo}%
\providecommand \bibfield  [0]{\@secondoftwo}%
\providecommand \translation [1]{[#1]}%
\providecommand \BibitemOpen [0]{}%
\providecommand \bibitemStop [0]{}%
\providecommand \bibitemNoStop [0]{.\EOS\space}%
\providecommand \EOS [0]{\spacefactor3000\relax}%
\providecommand \BibitemShut  [1]{\csname bibitem#1\endcsname}%
\let\auto@bib@innerbib\@empty
\bibitem [{\citenamefont {Vollhardt}\ and\ \citenamefont
  {W\"olfle}(2013)}]{vollhardt}%
  \BibitemOpen
  \bibfield  {author} {\bibinfo {author} {\bibfnamefont {D.}~\bibnamefont
  {Vollhardt}}\ and\ \bibinfo {author} {\bibfnamefont {P.}~\bibnamefont
  {W\"olfle}},\ }\href@noop {} {\emph {\bibinfo {title} {The Superfluid Phases
  of Helium 3}}},\ Dover Books on Physics Series\ (\bibinfo  {publisher} {Dover
  Publications},\ \bibinfo {year} {2013})\BibitemShut {NoStop}%
\bibitem [{\citenamefont {Volovik}(2009)}]{volovik}%
  \BibitemOpen
  \bibfield  {author} {\bibinfo {author} {\bibfnamefont {G.}~\bibnamefont
  {Volovik}},\ }\href@noop {} {\emph {\bibinfo {title} {The Universe in a
  Helium Droplet}}},\ International Series of Monographs on Physics\ (\bibinfo
  {publisher} {OUP Oxford},\ \bibinfo {year} {2009})\BibitemShut {NoStop}%
\bibitem [{\citenamefont {Jérome}\ \emph {et~al.}(1980)\citenamefont {Jérome},
  \citenamefont {Mazaud}, \citenamefont {Ribault},\ and\ \citenamefont
  {Bechgaard}}]{jerome}%
  \BibitemOpen
  \bibfield  {author} {\bibinfo {author} {\bibfnamefont {D.}~\bibnamefont
  {Jérome}}, \bibinfo {author} {\bibfnamefont {A.}~\bibnamefont {Mazaud}},
  \bibinfo {author} {\bibfnamefont {M.}~\bibnamefont {Ribault}}, \ and\
  \bibinfo {author} {\bibfnamefont {K.}~\bibnamefont {Bechgaard}},\ }\href
  {\doibase 10.1051/jphyslet:0198000410409500} {\bibfield  {journal} {\bibinfo
  {journal} {J. Physique Lett.}\ }\textbf {\bibinfo {volume} {41}},\ \bibinfo
  {pages} {95} (\bibinfo {year} {1980})}\BibitemShut {NoStop}%
\bibitem [{\citenamefont {Ishida}\ \emph {et~al.}(1998)\citenamefont {Ishida},
  \citenamefont {Mukuda}, \citenamefont {Kitaoka}, \citenamefont {Asayama},
  \citenamefont {Mao}, \citenamefont {Mori},\ and\ \citenamefont
  {Maeno}}]{SrRuO}%
  \BibitemOpen
  \bibfield  {author} {\bibinfo {author} {\bibfnamefont {K.}~\bibnamefont
  {Ishida}}, \bibinfo {author} {\bibfnamefont {H.}~\bibnamefont {Mukuda}},
  \bibinfo {author} {\bibfnamefont {Y.}~\bibnamefont {Kitaoka}}, \bibinfo
  {author} {\bibfnamefont {K.}~\bibnamefont {Asayama}}, \bibinfo {author}
  {\bibfnamefont {Z.~Q.}\ \bibnamefont {Mao}}, \bibinfo {author} {\bibfnamefont
  {Y.}~\bibnamefont {Mori}}, \ and\ \bibinfo {author} {\bibfnamefont
  {Y.}~\bibnamefont {Maeno}},\ }\href {\doibase 10.1038/25315} {\bibfield
  {journal} {\bibinfo  {journal} {Nature}\ }\textbf {\bibinfo {volume} {396}},\
  \bibinfo {pages} {658} (\bibinfo {year} {1998})}\BibitemShut {NoStop}%
\bibitem [{\citenamefont {Mackenzie}\ and\ \citenamefont
  {Maeno}(2003)}]{Mackenzie_RevModPhys.75.657}%
  \BibitemOpen
  \bibfield  {author} {\bibinfo {author} {\bibfnamefont {A.~P.}\ \bibnamefont
  {Mackenzie}}\ and\ \bibinfo {author} {\bibfnamefont {Y.}~\bibnamefont
  {Maeno}},\ }\href {\doibase 10.1103/RevModPhys.75.657} {\bibfield  {journal}
  {\bibinfo  {journal} {Rev. Mod. Phys.}\ }\textbf {\bibinfo {volume} {75}},\
  \bibinfo {pages} {657} (\bibinfo {year} {2003})}\BibitemShut {NoStop}%
\bibitem [{\citenamefont {Lebed}(2008)}]{lebed}%
  \BibitemOpen
  \bibfield  {author} {\bibinfo {author} {\bibfnamefont {A.}~\bibnamefont
  {Lebed}},\ }\href@noop {} {\emph {\bibinfo {title} {The Physics of Organic
  Superconductors and Conductors}}},\ Springer Series in Materials Science\
  (\bibinfo  {publisher} {Springer Berlin Heidelberg},\ \bibinfo {year}
  {2008})\BibitemShut {NoStop}%
\bibitem [{\citenamefont {Mineev}\ and\ \citenamefont
  {Samokhin}(1999)}]{Mineev}%
  \BibitemOpen
  \bibfield  {author} {\bibinfo {author} {\bibfnamefont {V.}~\bibnamefont
  {Mineev}}\ and\ \bibinfo {author} {\bibfnamefont {K.~V.}\ \bibnamefont
  {Samokhin}},\ }\href@noop {} {\emph {\bibinfo {title} {Introduction to
  Unconventional Superconductivity}}}\ (\bibinfo  {publisher} {Taylor \&
  Francis},\ \bibinfo {year} {1999})\BibitemShut {NoStop}%
\bibitem [{\citenamefont {Larkin}\ and\ \citenamefont
  {Ovchinnikov}(1965)}]{Larkin_Ovchinnikov_1965}%
  \BibitemOpen
  \bibfield  {author} {\bibinfo {author} {\bibfnamefont {A.~I.}\ \bibnamefont
  {Larkin}}\ and\ \bibinfo {author} {\bibfnamefont {Y.~N.}\ \bibnamefont
  {Ovchinnikov}},\ }\href@noop {} {\bibfield  {journal} {\bibinfo  {journal}
  {Sov.~Phys.~JETP}\ }\textbf {\bibinfo {volume} {20}},\ \bibinfo {pages} {762}
  (\bibinfo {year} {1965})}\BibitemShut {NoStop}%
\bibitem [{\citenamefont {Fulde}\ and\ \citenamefont
  {Ferrell}(1964)}]{Fulde_Ferrell_1964}%
  \BibitemOpen
  \bibfield  {author} {\bibinfo {author} {\bibfnamefont {P.}~\bibnamefont
  {Fulde}}\ and\ \bibinfo {author} {\bibfnamefont {R.~A.}\ \bibnamefont
  {Ferrell}},\ }\href {\doibase 10.1103/PhysRev.135.A550} {\bibfield  {journal}
  {\bibinfo  {journal} {Phys.~Rev.}\ }\textbf {\bibinfo {volume} {135}},\
  \bibinfo {pages} {A550} (\bibinfo {year} {1964})}\BibitemShut {NoStop}%
\bibitem [{\citenamefont {Fulde}(1973)}]{Fulde_1973}%
  \BibitemOpen
  \bibfield  {author} {\bibinfo {author} {\bibfnamefont {P.}~\bibnamefont
  {Fulde}},\ }\href {\doibase 10.1080/00018737300101369} {\bibfield  {journal}
  {\bibinfo  {journal} {Advances in Physics}\ }\textbf {\bibinfo {volume}
  {22}},\ \bibinfo {pages} {667} (\bibinfo {year} {1973})}\BibitemShut
  {NoStop}%
\bibitem [{\citenamefont {Meservey}\ and\ \citenamefont
  {Tedrow}(1994)}]{Meservey_1994}%
  \BibitemOpen
  \bibfield  {author} {\bibinfo {author} {\bibfnamefont {R.}~\bibnamefont
  {Meservey}}\ and\ \bibinfo {author} {\bibfnamefont {P.}~\bibnamefont
  {Tedrow}},\ }\href {\doibase http://dx.doi.org/10.1016/0370-1573(94)90105-8}
  {\bibfield  {journal} {\bibinfo  {journal} {Physics Reports}\ }\textbf
  {\bibinfo {volume} {238}},\ \bibinfo {pages} {173 } (\bibinfo {year}
  {1994})}\BibitemShut {NoStop}%
\bibitem [{\citenamefont {Bulaevskii}\ \emph {et~al.}(1985)\citenamefont
  {Bulaevskii}, \citenamefont {Buzdin}, \citenamefont {Kulic},\ and\
  \citenamefont {Panjukov}}]{Bulaevskii_et_al_Adv_Phys}%
  \BibitemOpen
  \bibfield  {author} {\bibinfo {author} {\bibfnamefont {L.}~\bibnamefont
  {Bulaevskii}}, \bibinfo {author} {\bibfnamefont {A.}~\bibnamefont {Buzdin}},
  \bibinfo {author} {\bibfnamefont {M.}~\bibnamefont {Kulic}}, \ and\ \bibinfo
  {author} {\bibfnamefont {S.}~\bibnamefont {Panjukov}},\ }\href {\doibase
  10.1080/00018738500101741} {\bibfield  {journal} {\bibinfo  {journal}
  {Advances in Physics}\ }\textbf {\bibinfo {volume} {34}},\ \bibinfo {pages}
  {175} (\bibinfo {year} {1985})}\BibitemShut {NoStop}%
\bibitem [{\citenamefont {Bergeret}\ \emph {et~al.}(2005)\citenamefont
  {Bergeret}, \citenamefont {Volkov},\ and\ \citenamefont {Efetov}}]{BVErmp}%
  \BibitemOpen
  \bibfield  {author} {\bibinfo {author} {\bibfnamefont {F.~S.}\ \bibnamefont
  {Bergeret}}, \bibinfo {author} {\bibfnamefont {A.~F.}\ \bibnamefont
  {Volkov}}, \ and\ \bibinfo {author} {\bibfnamefont {K.~B.}\ \bibnamefont
  {Efetov}},\ }\href {\doibase 10.1103/RevModPhys.77.1321} {\bibfield
  {journal} {\bibinfo  {journal} {Rev. Mod. Phys.}\ }\textbf {\bibinfo {volume}
  {77}},\ \bibinfo {pages} {1321} (\bibinfo {year} {2005})}\BibitemShut
  {NoStop}%
\bibitem [{\citenamefont {Eschrig}(2011)}]{Eschrig_Ph_Today}%
  \BibitemOpen
  \bibfield  {author} {\bibinfo {author} {\bibfnamefont {M.}~\bibnamefont
  {Eschrig}},\ }\href {\doibase 10.1063/1.3541944} {\bibfield  {journal}
  {\bibinfo  {journal} {Physics Today}\ }\textbf {\bibinfo {volume} {64}},\
  \bibinfo {pages} {43} (\bibinfo {year} {2011})}\BibitemShut {NoStop}%
\bibitem [{\citenamefont {Buzdin}(2005)}]{Buzdin_RMP_2005}%
  \BibitemOpen
  \bibfield  {author} {\bibinfo {author} {\bibfnamefont {A.~I.}\ \bibnamefont
  {Buzdin}},\ }\href {\doibase 10.1103/RevModPhys.77.935} {\bibfield  {journal}
  {\bibinfo  {journal} {Rev. Mod. Phys.}\ }\textbf {\bibinfo {volume} {77}},\
  \bibinfo {pages} {935} (\bibinfo {year} {2005})}\BibitemShut {NoStop}%
\bibitem [{\citenamefont {Bergeret}\ \emph
  {et~al.}(2001{\natexlab{a}})\citenamefont {Bergeret}, \citenamefont
  {Volkov},\ and\ \citenamefont {Efetov}}]{Bergeret_Volkov_Efetov_2001}%
  \BibitemOpen
  \bibfield  {author} {\bibinfo {author} {\bibfnamefont {F.~S.}\ \bibnamefont
  {Bergeret}}, \bibinfo {author} {\bibfnamefont {A.~F.}\ \bibnamefont
  {Volkov}}, \ and\ \bibinfo {author} {\bibfnamefont {K.~B.}\ \bibnamefont
  {Efetov}},\ }\href {\doibase 10.1103/PhysRevLett.86.4096} {\bibfield
  {journal} {\bibinfo  {journal} {Phys. Rev. Lett.}\ }\textbf {\bibinfo
  {volume} {86}},\ \bibinfo {pages} {4096} (\bibinfo {year}
  {2001}{\natexlab{a}})}\BibitemShut {NoStop}%
\bibitem [{\citenamefont {Kadigrobov}\ \emph {et~al.}(2001)\citenamefont
  {Kadigrobov}, \citenamefont {Shekhter},\ and\ \citenamefont
  {Jonson}}]{Kad01}%
  \BibitemOpen
  \bibfield  {author} {\bibinfo {author} {\bibfnamefont {A.}~\bibnamefont
  {Kadigrobov}}, \bibinfo {author} {\bibfnamefont {R.~I.}\ \bibnamefont
  {Shekhter}}, \ and\ \bibinfo {author} {\bibfnamefont {M.}~\bibnamefont
  {Jonson}},\ }\href {http://stacks.iop.org/0295-5075/54/i=3/a=394} {\bibfield
  {journal} {\bibinfo  {journal} {EPL (Europhysics Letters)}\ }\textbf
  {\bibinfo {volume} {54}},\ \bibinfo {pages} {394} (\bibinfo {year}
  {2001})}\BibitemShut {NoStop}%
\bibitem [{\citenamefont {Volkov}\ and\ \citenamefont {Efetov}(2008)}]{VE08}%
  \BibitemOpen
  \bibfield  {author} {\bibinfo {author} {\bibfnamefont {A.~F.}\ \bibnamefont
  {Volkov}}\ and\ \bibinfo {author} {\bibfnamefont {K.~B.}\ \bibnamefont
  {Efetov}},\ }\href {\doibase 10.1103/PhysRevB.78.024519} {\bibfield
  {journal} {\bibinfo  {journal} {Phys. Rev. B}\ }\textbf {\bibinfo {volume}
  {78}},\ \bibinfo {pages} {024519} (\bibinfo {year} {2008})}\BibitemShut
  {NoStop}%
\bibitem [{\citenamefont {Asano}\ \emph {et~al.}(2007)\citenamefont {Asano},
  \citenamefont {Sawa}, \citenamefont {Tanaka},\ and\ \citenamefont
  {Golubov}}]{Asano_et_al_2007}%
  \BibitemOpen
  \bibfield  {author} {\bibinfo {author} {\bibfnamefont {Y.}~\bibnamefont
  {Asano}}, \bibinfo {author} {\bibfnamefont {Y.}~\bibnamefont {Sawa}},
  \bibinfo {author} {\bibfnamefont {Y.}~\bibnamefont {Tanaka}}, \ and\ \bibinfo
  {author} {\bibfnamefont {A.~A.}\ \bibnamefont {Golubov}},\ }\href {\doibase
  10.1103/PhysRevB.76.224525} {\bibfield  {journal} {\bibinfo  {journal} {Phys.
  Rev. B}\ }\textbf {\bibinfo {volume} {76}},\ \bibinfo {pages} {224525}
  (\bibinfo {year} {2007})}\BibitemShut {NoStop}%
\bibitem [{\citenamefont {Houzet}\ and\ \citenamefont
  {Buzdin}(2007)}]{Buzdin07}%
  \BibitemOpen
  \bibfield  {author} {\bibinfo {author} {\bibfnamefont {M.}~\bibnamefont
  {Houzet}}\ and\ \bibinfo {author} {\bibfnamefont {A.~I.}\ \bibnamefont
  {Buzdin}},\ }\href {\doibase 10.1103/PhysRevB.76.060504} {\bibfield
  {journal} {\bibinfo  {journal} {Phys. Rev. B}\ }\textbf {\bibinfo {volume}
  {76}},\ \bibinfo {pages} {060504(R)} (\bibinfo {year} {2007})}\BibitemShut
  {NoStop}%
\bibitem [{\citenamefont {Galaktionov}\ \emph {et~al.}(2008)\citenamefont
  {Galaktionov}, \citenamefont {Kalenkov},\ and\ \citenamefont
  {Zaikin}}]{Zaikin08}%
  \BibitemOpen
  \bibfield  {author} {\bibinfo {author} {\bibfnamefont {A.~V.}\ \bibnamefont
  {Galaktionov}}, \bibinfo {author} {\bibfnamefont {M.~S.}\ \bibnamefont
  {Kalenkov}}, \ and\ \bibinfo {author} {\bibfnamefont {A.~D.}\ \bibnamefont
  {Zaikin}},\ }\href {\doibase 10.1103/PhysRevB.77.094520} {\bibfield
  {journal} {\bibinfo  {journal} {Phys. Rev. B}\ }\textbf {\bibinfo {volume}
  {77}},\ \bibinfo {pages} {094520} (\bibinfo {year} {2008})}\BibitemShut
  {NoStop}%
\bibitem [{\citenamefont {Linder}\ \emph {et~al.}(2009)\citenamefont {Linder},
  \citenamefont {Yokoyama},\ and\ \citenamefont {Sudb\o{}}}]{Linder09}%
  \BibitemOpen
  \bibfield  {author} {\bibinfo {author} {\bibfnamefont {J.}~\bibnamefont
  {Linder}}, \bibinfo {author} {\bibfnamefont {T.}~\bibnamefont {Yokoyama}}, \
  and\ \bibinfo {author} {\bibfnamefont {A.}~\bibnamefont {Sudb\o{}}},\ }\href
  {\doibase 10.1103/PhysRevB.79.054523} {\bibfield  {journal} {\bibinfo
  {journal} {Phys. Rev. B}\ }\textbf {\bibinfo {volume} {79}},\ \bibinfo
  {pages} {054523} (\bibinfo {year} {2009})}\BibitemShut {NoStop}%
\bibitem [{\citenamefont {Trifunovic}\ and\ \citenamefont
  {Radovi\ifmmode~\acute{c}\else \'{c}\fi{}}(2010)}]{Radovic10}%
  \BibitemOpen
  \bibfield  {author} {\bibinfo {author} {\bibfnamefont {L.}~\bibnamefont
  {Trifunovic}}\ and\ \bibinfo {author} {\bibfnamefont {Z.}~\bibnamefont
  {Radovi\ifmmode~\acute{c}\else \'{c}\fi{}}},\ }\href {\doibase
  10.1103/PhysRevB.82.020505} {\bibfield  {journal} {\bibinfo  {journal} {Phys.
  Rev. B}\ }\textbf {\bibinfo {volume} {82}},\ \bibinfo {pages} {020505}
  (\bibinfo {year} {2010})}\BibitemShut {NoStop}%
\bibitem [{\citenamefont {Halterman}\ \emph {et~al.}(2008)\citenamefont
  {Halterman}, \citenamefont {Valls},\ and\ \citenamefont {Barsic}}]{Valls08}%
  \BibitemOpen
  \bibfield  {author} {\bibinfo {author} {\bibfnamefont {K.}~\bibnamefont
  {Halterman}}, \bibinfo {author} {\bibfnamefont {O.~T.}\ \bibnamefont
  {Valls}}, \ and\ \bibinfo {author} {\bibfnamefont {P.~H.}\ \bibnamefont
  {Barsic}},\ }\href {\doibase 10.1103/PhysRevB.77.174511} {\bibfield
  {journal} {\bibinfo  {journal} {Phys. Rev. B}\ }\textbf {\bibinfo {volume}
  {77}},\ \bibinfo {pages} {174511} (\bibinfo {year} {2008})}\BibitemShut
  {NoStop}%
\bibitem [{\citenamefont {{Halterman}}\ \emph {et~al.}(2015)\citenamefont
  {{Halterman}}, \citenamefont {{Valls}},\ and\ \citenamefont
  {{Wu}}}]{Halterman_et_al_arXiv_2015}%
  \BibitemOpen
  \bibfield  {author} {\bibinfo {author} {\bibfnamefont {K.}~\bibnamefont
  {{Halterman}}}, \bibinfo {author} {\bibfnamefont {O.~T.}\ \bibnamefont
  {{Valls}}}, \ and\ \bibinfo {author} {\bibfnamefont {C.-T.}\ \bibnamefont
  {{Wu}}},\ }\href@noop {} {\bibfield  {journal} {\bibinfo  {journal} {ArXiv
  e-prints}\ } (\bibinfo {year} {2015})},\ \Eprint
  {http://arxiv.org/abs/1506.05489} {arXiv:1506.05489 [cond-mat.supr-con]}
  \BibitemShut {NoStop}%
\bibitem [{\citenamefont {Alidoust}\ \emph {et~al.}(2015)\citenamefont
  {Alidoust}, \citenamefont {Halterman},\ and\ \citenamefont
  {Valls}}]{Alidoust_et_al_2015}%
  \BibitemOpen
  \bibfield  {author} {\bibinfo {author} {\bibfnamefont {M.}~\bibnamefont
  {Alidoust}}, \bibinfo {author} {\bibfnamefont {K.}~\bibnamefont {Halterman}},
  \ and\ \bibinfo {author} {\bibfnamefont {O.~T.}\ \bibnamefont {Valls}},\
  }\href {\doibase 10.1103/PhysRevB.92.014508} {\bibfield  {journal} {\bibinfo
  {journal} {Phys. Rev. B}\ }\textbf {\bibinfo {volume} {92}},\ \bibinfo
  {pages} {014508} (\bibinfo {year} {2015})}\BibitemShut {NoStop}%
\bibitem [{\citenamefont {{Leksin}}\ \emph {et~al.}(2015)\citenamefont
  {{Leksin}}, \citenamefont {{Garifyanov}}, \citenamefont {{Kamashev}},
  \citenamefont {{Fominov}}, \citenamefont {{Schumann}}, \citenamefont
  {{Hess}}, \citenamefont {{Kataev}}, \citenamefont {{Buechner}},\ and\
  \citenamefont {{Garifullin}}}]{Leksin_et_al_arXiv_2015}%
  \BibitemOpen
  \bibfield  {author} {\bibinfo {author} {\bibfnamefont {P.~V.}\ \bibnamefont
  {{Leksin}}}, \bibinfo {author} {\bibfnamefont {N.~N.}\ \bibnamefont
  {{Garifyanov}}}, \bibinfo {author} {\bibfnamefont {A.~A.}\ \bibnamefont
  {{Kamashev}}}, \bibinfo {author} {\bibfnamefont {Y.~V.}\ \bibnamefont
  {{Fominov}}}, \bibinfo {author} {\bibfnamefont {J.}~\bibnamefont
  {{Schumann}}}, \bibinfo {author} {\bibfnamefont {C.}~\bibnamefont {{Hess}}},
  \bibinfo {author} {\bibfnamefont {V.}~\bibnamefont {{Kataev}}}, \bibinfo
  {author} {\bibfnamefont {B.}~\bibnamefont {{Buechner}}}, \ and\ \bibinfo
  {author} {\bibfnamefont {I.~A.}\ \bibnamefont {{Garifullin}}},\ }\href@noop
  {} {\bibfield  {journal} {\bibinfo  {journal} {ArXiv e-prints}\ } (\bibinfo
  {year} {2015})},\ \Eprint {http://arxiv.org/abs/1505.07849} {arXiv:1505.07849
  [cond-mat.supr-con]} \BibitemShut {NoStop}%
\bibitem [{\citenamefont {Fominov}\ \emph {et~al.}(2015)\citenamefont
  {Fominov}, \citenamefont {Tanaka}, \citenamefont {Asano},\ and\ \citenamefont
  {Eschrig}}]{Fominov_et_al_2015}%
  \BibitemOpen
  \bibfield  {author} {\bibinfo {author} {\bibfnamefont {Y.~V.}\ \bibnamefont
  {Fominov}}, \bibinfo {author} {\bibfnamefont {Y.}~\bibnamefont {Tanaka}},
  \bibinfo {author} {\bibfnamefont {Y.}~\bibnamefont {Asano}}, \ and\ \bibinfo
  {author} {\bibfnamefont {M.}~\bibnamefont {Eschrig}},\ }\href {\doibase
  10.1103/PhysRevB.91.144514} {\bibfield  {journal} {\bibinfo  {journal} {Phys.
  Rev. B}\ }\textbf {\bibinfo {volume} {91}},\ \bibinfo {pages} {144514}
  (\bibinfo {year} {2015})}\BibitemShut {NoStop}%
\bibitem [{\citenamefont {Banerjee}\ \emph {et~al.}(2014)\citenamefont
  {Banerjee}, \citenamefont {Robinson},\ and\ \citenamefont
  {Blamire}}]{Banerjee_et_al_2014}%
  \BibitemOpen
  \bibfield  {author} {\bibinfo {author} {\bibfnamefont {N.}~\bibnamefont
  {Banerjee}}, \bibinfo {author} {\bibfnamefont {J.~W.~A.}\ \bibnamefont
  {Robinson}}, \ and\ \bibinfo {author} {\bibfnamefont {M.~G.}\ \bibnamefont
  {Blamire}},\ }\href {\doibase 10.1038/ncomms5771} {\bibfield  {journal}
  {\bibinfo  {journal} {Nat. Commun.}\ }\textbf {\bibinfo {volume} {5}},\
  \bibinfo {pages} {4771} (\bibinfo {year} {2014})}\BibitemShut {NoStop}%
\bibitem [{\citenamefont {Keizer}\ \emph {et~al.}(2006)\citenamefont {Keizer},
  \citenamefont {Goennenwein}, \citenamefont {Klapwijk}, \citenamefont {Miao},
  \citenamefont {Xiao},\ and\ \citenamefont {Gupta}}]{Keizer06}%
  \BibitemOpen
  \bibfield  {author} {\bibinfo {author} {\bibfnamefont {R.~S.}\ \bibnamefont
  {Keizer}}, \bibinfo {author} {\bibfnamefont {S.~T.~B.}\ \bibnamefont
  {Goennenwein}}, \bibinfo {author} {\bibfnamefont {T.~M.}\ \bibnamefont
  {Klapwijk}}, \bibinfo {author} {\bibfnamefont {G.}~\bibnamefont {Miao}},
  \bibinfo {author} {\bibfnamefont {G.}~\bibnamefont {Xiao}}, \ and\ \bibinfo
  {author} {\bibfnamefont {A.}~\bibnamefont {Gupta}},\ }\href {\doibase
  10.1038/nature04499} {\bibfield  {journal} {\bibinfo  {journal} {Nature}\
  }\textbf {\bibinfo {volume} {439}},\ \bibinfo {pages} {825} (\bibinfo {year}
  {2006})}\BibitemShut {NoStop}%
\bibitem [{\citenamefont {Sosnin}\ \emph {et~al.}(2006)\citenamefont {Sosnin},
  \citenamefont {Cho}, \citenamefont {Petrashov},\ and\ \citenamefont
  {Volkov}}]{Petrashov06}%
  \BibitemOpen
  \bibfield  {author} {\bibinfo {author} {\bibfnamefont {I.}~\bibnamefont
  {Sosnin}}, \bibinfo {author} {\bibfnamefont {H.}~\bibnamefont {Cho}},
  \bibinfo {author} {\bibfnamefont {V.~T.}\ \bibnamefont {Petrashov}}, \ and\
  \bibinfo {author} {\bibfnamefont {A.~F.}\ \bibnamefont {Volkov}},\ }\href
  {\doibase 10.1103/PhysRevLett.96.157002} {\bibfield  {journal} {\bibinfo
  {journal} {Phys. Rev. Lett.}\ }\textbf {\bibinfo {volume} {96}},\ \bibinfo
  {pages} {157002} (\bibinfo {year} {2006})}\BibitemShut {NoStop}%
\bibitem [{\citenamefont {Khaire}\ \emph {et~al.}(2010)\citenamefont {Khaire},
  \citenamefont {Khasawneh}, \citenamefont {Pratt},\ and\ \citenamefont
  {Birge}}]{Birge10}%
  \BibitemOpen
  \bibfield  {author} {\bibinfo {author} {\bibfnamefont {T.~S.}\ \bibnamefont
  {Khaire}}, \bibinfo {author} {\bibfnamefont {M.~A.}\ \bibnamefont
  {Khasawneh}}, \bibinfo {author} {\bibfnamefont {W.~P.}\ \bibnamefont
  {Pratt}}, \ and\ \bibinfo {author} {\bibfnamefont {N.~O.}\ \bibnamefont
  {Birge}},\ }\href {\doibase 10.1103/PhysRevLett.104.137002} {\bibfield
  {journal} {\bibinfo  {journal} {Phys. Rev. Lett.}\ }\textbf {\bibinfo
  {volume} {104}},\ \bibinfo {pages} {137002} (\bibinfo {year}
  {2010})}\BibitemShut {NoStop}%
\bibitem [{\citenamefont {Klose}\ \emph {et~al.}(2012)\citenamefont {Klose},
  \citenamefont {Khaire}, \citenamefont {Wang}, \citenamefont {Pratt},
  \citenamefont {Birge}, \citenamefont {McMorran}, \citenamefont {Ginley},
  \citenamefont {Borchers}, \citenamefont {Kirby}, \citenamefont {Maranville},\
  and\ \citenamefont {Unguris}}]{Birge12}%
  \BibitemOpen
  \bibfield  {author} {\bibinfo {author} {\bibfnamefont {C.}~\bibnamefont
  {Klose}}, \bibinfo {author} {\bibfnamefont {T.~S.}\ \bibnamefont {Khaire}},
  \bibinfo {author} {\bibfnamefont {Y.}~\bibnamefont {Wang}}, \bibinfo {author}
  {\bibfnamefont {W.~P.}\ \bibnamefont {Pratt}}, \bibinfo {author}
  {\bibfnamefont {N.~O.}\ \bibnamefont {Birge}}, \bibinfo {author}
  {\bibfnamefont {B.~J.}\ \bibnamefont {McMorran}}, \bibinfo {author}
  {\bibfnamefont {T.~P.}\ \bibnamefont {Ginley}}, \bibinfo {author}
  {\bibfnamefont {J.~A.}\ \bibnamefont {Borchers}}, \bibinfo {author}
  {\bibfnamefont {B.~J.}\ \bibnamefont {Kirby}}, \bibinfo {author}
  {\bibfnamefont {B.~B.}\ \bibnamefont {Maranville}}, \ and\ \bibinfo {author}
  {\bibfnamefont {J.}~\bibnamefont {Unguris}},\ }\href {\doibase
  10.1103/PhysRevLett.108.127002} {\bibfield  {journal} {\bibinfo  {journal}
  {Phys. Rev. Lett.}\ }\textbf {\bibinfo {volume} {108}},\ \bibinfo {pages}
  {127002} (\bibinfo {year} {2012})}\BibitemShut {NoStop}%
\bibitem [{\citenamefont {Anwar}\ \emph {et~al.}(2010)\citenamefont {Anwar},
  \citenamefont {Czeschka}, \citenamefont {Hesselberth}, \citenamefont
  {Porcu},\ and\ \citenamefont {Aarts}}]{Aarts10}%
  \BibitemOpen
  \bibfield  {author} {\bibinfo {author} {\bibfnamefont {M.~S.}\ \bibnamefont
  {Anwar}}, \bibinfo {author} {\bibfnamefont {F.}~\bibnamefont {Czeschka}},
  \bibinfo {author} {\bibfnamefont {M.}~\bibnamefont {Hesselberth}}, \bibinfo
  {author} {\bibfnamefont {M.}~\bibnamefont {Porcu}}, \ and\ \bibinfo {author}
  {\bibfnamefont {J.}~\bibnamefont {Aarts}},\ }\href {\doibase
  10.1103/PhysRevB.82.100501} {\bibfield  {journal} {\bibinfo  {journal} {Phys.
  Rev. B}\ }\textbf {\bibinfo {volume} {82}},\ \bibinfo {pages} {100501}
  (\bibinfo {year} {2010})}\BibitemShut {NoStop}%
\bibitem [{\citenamefont {Anwar}\ and\ \citenamefont {Aarts}(2011)}]{Aarts11}%
  \BibitemOpen
  \bibfield  {author} {\bibinfo {author} {\bibfnamefont {M.~S.}\ \bibnamefont
  {Anwar}}\ and\ \bibinfo {author} {\bibfnamefont {J.}~\bibnamefont {Aarts}},\
  }\href {http://stacks.iop.org/0953-2048/24/i=2/a=024016} {\bibfield
  {journal} {\bibinfo  {journal} {Superconductor Science and Technology}\
  }\textbf {\bibinfo {volume} {24}},\ \bibinfo {pages} {024016} (\bibinfo
  {year} {2011})}\BibitemShut {NoStop}%
\bibitem [{\citenamefont {Anwar}\ \emph {et~al.}(2012)\citenamefont {Anwar},
  \citenamefont {Veldhorst}, \citenamefont {Brinkman},\ and\ \citenamefont
  {Aarts}}]{Aarts12}%
  \BibitemOpen
  \bibfield  {author} {\bibinfo {author} {\bibfnamefont {M.~S.}\ \bibnamefont
  {Anwar}}, \bibinfo {author} {\bibfnamefont {M.}~\bibnamefont {Veldhorst}},
  \bibinfo {author} {\bibfnamefont {A.}~\bibnamefont {Brinkman}}, \ and\
  \bibinfo {author} {\bibfnamefont {J.}~\bibnamefont {Aarts}},\ }\href
  {\doibase http://dx.doi.org/10.1063/1.3681138} {\bibfield  {journal}
  {\bibinfo  {journal} {Applied Physics Letters}\ }\textbf {\bibinfo {volume}
  {100}},\ \bibinfo {eid} {052602} (\bibinfo {year} {2012})}\BibitemShut
  {NoStop}%
\bibitem [{\citenamefont {Robinson}\ \emph {et~al.}(2010)\citenamefont
  {Robinson}, \citenamefont {Witt},\ and\ \citenamefont {Blamire}}]{Blamire10}%
  \BibitemOpen
  \bibfield  {author} {\bibinfo {author} {\bibfnamefont {J.~W.~A.}\
  \bibnamefont {Robinson}}, \bibinfo {author} {\bibfnamefont {J.~D.~S.}\
  \bibnamefont {Witt}}, \ and\ \bibinfo {author} {\bibfnamefont {M.~G.}\
  \bibnamefont {Blamire}},\ }\href {\doibase 10.1126/science.1189246}
  {\bibfield  {journal} {\bibinfo  {journal} {Science}\ }\textbf {\bibinfo
  {volume} {329}},\ \bibinfo {pages} {59} (\bibinfo {year} {2010})}\BibitemShut
  {NoStop}%
\bibitem [{\citenamefont {Witt}\ \emph {et~al.}(2012)\citenamefont {Witt},
  \citenamefont {Robinson},\ and\ \citenamefont {Blamire}}]{Blamire12}%
  \BibitemOpen
  \bibfield  {author} {\bibinfo {author} {\bibfnamefont {J.~D.~S.}\
  \bibnamefont {Witt}}, \bibinfo {author} {\bibfnamefont {J.~W.~A.}\
  \bibnamefont {Robinson}}, \ and\ \bibinfo {author} {\bibfnamefont {M.~G.}\
  \bibnamefont {Blamire}},\ }\href {\doibase 10.1103/PhysRevB.85.184526}
  {\bibfield  {journal} {\bibinfo  {journal} {Phys. Rev. B}\ }\textbf {\bibinfo
  {volume} {85}},\ \bibinfo {pages} {184526} (\bibinfo {year}
  {2012})}\BibitemShut {NoStop}%
\bibitem [{\citenamefont {Sprungmann}\ \emph {et~al.}(2010)\citenamefont
  {Sprungmann}, \citenamefont {Westerholt}, \citenamefont {Zabel},
  \citenamefont {Weides},\ and\ \citenamefont {Kohlstedt}}]{Zabel10}%
  \BibitemOpen
  \bibfield  {author} {\bibinfo {author} {\bibfnamefont {D.}~\bibnamefont
  {Sprungmann}}, \bibinfo {author} {\bibfnamefont {K.}~\bibnamefont
  {Westerholt}}, \bibinfo {author} {\bibfnamefont {H.}~\bibnamefont {Zabel}},
  \bibinfo {author} {\bibfnamefont {M.}~\bibnamefont {Weides}}, \ and\ \bibinfo
  {author} {\bibfnamefont {H.}~\bibnamefont {Kohlstedt}},\ }\href {\doibase
  10.1103/PhysRevB.82.060505} {\bibfield  {journal} {\bibinfo  {journal} {Phys.
  Rev. B}\ }\textbf {\bibinfo {volume} {82}},\ \bibinfo {pages} {060505}
  (\bibinfo {year} {2010})}\BibitemShut {NoStop}%
\bibitem [{\citenamefont {Kalenkov}\ \emph {et~al.}(2011)\citenamefont
  {Kalenkov}, \citenamefont {Zaikin},\ and\ \citenamefont
  {Petrashov}}]{Petrashov11}%
  \BibitemOpen
  \bibfield  {author} {\bibinfo {author} {\bibfnamefont {M.~S.}\ \bibnamefont
  {Kalenkov}}, \bibinfo {author} {\bibfnamefont {A.~D.}\ \bibnamefont
  {Zaikin}}, \ and\ \bibinfo {author} {\bibfnamefont {V.~T.}\ \bibnamefont
  {Petrashov}},\ }\href {\doibase 10.1103/PhysRevLett.107.087003} {\bibfield
  {journal} {\bibinfo  {journal} {Phys. Rev. Lett.}\ }\textbf {\bibinfo
  {volume} {107}},\ \bibinfo {pages} {087003} (\bibinfo {year}
  {2011})}\BibitemShut {NoStop}%
\bibitem [{\citenamefont {Kalcheim}\ \emph {et~al.}(2011)\citenamefont
  {Kalcheim}, \citenamefont {Kirzhner}, \citenamefont {Koren},\ and\
  \citenamefont {Millo}}]{Kalcheim_et_al_2011}%
  \BibitemOpen
  \bibfield  {author} {\bibinfo {author} {\bibfnamefont {Y.}~\bibnamefont
  {Kalcheim}}, \bibinfo {author} {\bibfnamefont {T.}~\bibnamefont {Kirzhner}},
  \bibinfo {author} {\bibfnamefont {G.}~\bibnamefont {Koren}}, \ and\ \bibinfo
  {author} {\bibfnamefont {O.}~\bibnamefont {Millo}},\ }\href {\doibase
  10.1103/PhysRevB.83.064510} {\bibfield  {journal} {\bibinfo  {journal} {Phys.
  Rev. B}\ }\textbf {\bibinfo {volume} {83}},\ \bibinfo {pages} {064510}
  (\bibinfo {year} {2011})}\BibitemShut {NoStop}%
\bibitem [{\citenamefont {Leksin}\ \emph {et~al.}(2012)\citenamefont {Leksin},
  \citenamefont {Garif'yanov}, \citenamefont {Garifullin}, \citenamefont
  {Fominov}, \citenamefont {Schumann}, \citenamefont {Krupskaya}, \citenamefont
  {Kataev}, \citenamefont {Schmidt},\ and\ \citenamefont
  {B\"uchner}}]{Leksin_et_al_2012}%
  \BibitemOpen
  \bibfield  {author} {\bibinfo {author} {\bibfnamefont {P.~V.}\ \bibnamefont
  {Leksin}}, \bibinfo {author} {\bibfnamefont {N.~N.}\ \bibnamefont
  {Garif'yanov}}, \bibinfo {author} {\bibfnamefont {I.~A.}\ \bibnamefont
  {Garifullin}}, \bibinfo {author} {\bibfnamefont {Y.~V.}\ \bibnamefont
  {Fominov}}, \bibinfo {author} {\bibfnamefont {J.}~\bibnamefont {Schumann}},
  \bibinfo {author} {\bibfnamefont {Y.}~\bibnamefont {Krupskaya}}, \bibinfo
  {author} {\bibfnamefont {V.}~\bibnamefont {Kataev}}, \bibinfo {author}
  {\bibfnamefont {O.~G.}\ \bibnamefont {Schmidt}}, \ and\ \bibinfo {author}
  {\bibfnamefont {B.}~\bibnamefont {B\"uchner}},\ }\href {\doibase
  10.1103/PhysRevLett.109.057005} {\bibfield  {journal} {\bibinfo  {journal}
  {Phys. Rev. Lett.}\ }\textbf {\bibinfo {volume} {109}},\ \bibinfo {pages}
  {057005} (\bibinfo {year} {2012})}\BibitemShut {NoStop}%
\bibitem [{\citenamefont {Khaydukov}\ \emph {et~al.}(2014)\citenamefont
  {Khaydukov}, \citenamefont {Ovsyannikov}, \citenamefont {Sheyerman},
  \citenamefont {Constantinian}, \citenamefont {Mustafa}, \citenamefont
  {Keller}, \citenamefont {Uribe-Laverde}, \citenamefont {Kislinskii},
  \citenamefont {Shadrin}, \citenamefont {Kalaboukhov}, \citenamefont
  {Keimer},\ and\ \citenamefont {Winkler}}]{Khaydukov_et_al_2014}%
  \BibitemOpen
  \bibfield  {author} {\bibinfo {author} {\bibfnamefont {Y.~N.}\ \bibnamefont
  {Khaydukov}}, \bibinfo {author} {\bibfnamefont {G.~A.}\ \bibnamefont
  {Ovsyannikov}}, \bibinfo {author} {\bibfnamefont {A.~E.}\ \bibnamefont
  {Sheyerman}}, \bibinfo {author} {\bibfnamefont {K.~Y.}\ \bibnamefont
  {Constantinian}}, \bibinfo {author} {\bibfnamefont {L.}~\bibnamefont
  {Mustafa}}, \bibinfo {author} {\bibfnamefont {T.}~\bibnamefont {Keller}},
  \bibinfo {author} {\bibfnamefont {M.~A.}\ \bibnamefont {Uribe-Laverde}},
  \bibinfo {author} {\bibfnamefont {Y.~V.}\ \bibnamefont {Kislinskii}},
  \bibinfo {author} {\bibfnamefont {A.~V.}\ \bibnamefont {Shadrin}}, \bibinfo
  {author} {\bibfnamefont {A.}~\bibnamefont {Kalaboukhov}}, \bibinfo {author}
  {\bibfnamefont {B.}~\bibnamefont {Keimer}}, \ and\ \bibinfo {author}
  {\bibfnamefont {D.}~\bibnamefont {Winkler}},\ }\href {\doibase
  10.1103/PhysRevB.90.035130} {\bibfield  {journal} {\bibinfo  {journal} {Phys.
  Rev. B}\ }\textbf {\bibinfo {volume} {90}},\ \bibinfo {pages} {035130}
  (\bibinfo {year} {2014})}\BibitemShut {NoStop}%
\bibitem [{\citenamefont {Bulaevskii}\ \emph {et~al.}(1980)\citenamefont
  {Bulaevskii}, \citenamefont {Rusinov},\ and\ \citenamefont
  {Kulic}}]{Bulaevskii_Rusinov_Kulic_1980}%
  \BibitemOpen
  \bibfield  {author} {\bibinfo {author} {\bibfnamefont {L.}~\bibnamefont
  {Bulaevskii}}, \bibinfo {author} {\bibfnamefont {A.}~\bibnamefont {Rusinov}},
  \ and\ \bibinfo {author} {\bibfnamefont {M.}~\bibnamefont {Kulic}},\ }\href
  {\doibase 10.1007/BF00115620} {\bibfield  {journal} {\bibinfo  {journal}
  {Journal of Low Temperature Physics}\ }\textbf {\bibinfo {volume} {39}},\
  \bibinfo {pages} {255} (\bibinfo {year} {1980})}\BibitemShut {NoStop}%
\bibitem [{\citenamefont {Volkov}(2008)}]{Volkov_2008}%
  \BibitemOpen
  \bibfield  {author} {\bibinfo {author} {\bibfnamefont {A.~F.}\ \bibnamefont
  {Volkov}},\ }\href {\doibase 10.1103/PhysRevB.77.064521} {\bibfield
  {journal} {\bibinfo  {journal} {Phys. Rev. B}\ }\textbf {\bibinfo {volume}
  {77}},\ \bibinfo {pages} {064521} (\bibinfo {year} {2008})}\BibitemShut
  {NoStop}%
\bibitem [{\citenamefont {Linder}\ and\ \citenamefont
  {Robinson}(2015)}]{Linder_Robinson_2015}%
  \BibitemOpen
  \bibfield  {author} {\bibinfo {author} {\bibfnamefont {J.}~\bibnamefont
  {Linder}}\ and\ \bibinfo {author} {\bibfnamefont {J.~W.~A.}\ \bibnamefont
  {Robinson}},\ }\href {\doibase 10.1038/nphys3242} {\bibfield  {journal}
  {\bibinfo  {journal} {Nat.~Phys.}\ }\textbf {\bibinfo {volume} {11}},\
  \bibinfo {pages} {307} (\bibinfo {year} {2015})}\BibitemShut {NoStop}%
\bibitem [{\citenamefont {Bergeret}\ \emph
  {et~al.}(2001{\natexlab{b}})\citenamefont {Bergeret}, \citenamefont
  {Volkov},\ and\ \citenamefont {Efetov}}]{Bergeret_Volkov_Efetov_2001_b}%
  \BibitemOpen
  \bibfield  {author} {\bibinfo {author} {\bibfnamefont {F.~S.}\ \bibnamefont
  {Bergeret}}, \bibinfo {author} {\bibfnamefont {A.~F.}\ \bibnamefont
  {Volkov}}, \ and\ \bibinfo {author} {\bibfnamefont {K.~B.}\ \bibnamefont
  {Efetov}},\ }\href {\doibase 10.1103/PhysRevLett.86.3140} {\bibfield
  {journal} {\bibinfo  {journal} {Phys. Rev. Lett.}\ }\textbf {\bibinfo
  {volume} {86}},\ \bibinfo {pages} {3140} (\bibinfo {year}
  {2001}{\natexlab{b}})}\BibitemShut {NoStop}%
\bibitem [{\citenamefont {Santos}\ \emph {et~al.}(2008)\citenamefont {Santos},
  \citenamefont {Moodera}, \citenamefont {Raman}, \citenamefont {Negusse},
  \citenamefont {Holroyd}, \citenamefont {Dvorak}, \citenamefont {Liberati},
  \citenamefont {Idzerda},\ and\ \citenamefont {Arenholz}}]{Moodera08}%
  \BibitemOpen
  \bibfield  {author} {\bibinfo {author} {\bibfnamefont {T.~S.}\ \bibnamefont
  {Santos}}, \bibinfo {author} {\bibfnamefont {J.~S.}\ \bibnamefont {Moodera}},
  \bibinfo {author} {\bibfnamefont {K.~V.}\ \bibnamefont {Raman}}, \bibinfo
  {author} {\bibfnamefont {E.}~\bibnamefont {Negusse}}, \bibinfo {author}
  {\bibfnamefont {J.}~\bibnamefont {Holroyd}}, \bibinfo {author} {\bibfnamefont
  {J.}~\bibnamefont {Dvorak}}, \bibinfo {author} {\bibfnamefont
  {M.}~\bibnamefont {Liberati}}, \bibinfo {author} {\bibfnamefont {Y.~U.}\
  \bibnamefont {Idzerda}}, \ and\ \bibinfo {author} {\bibfnamefont
  {E.}~\bibnamefont {Arenholz}},\ }\href {\doibase
  10.1103/PhysRevLett.101.147201} {\bibfield  {journal} {\bibinfo  {journal}
  {Phys. Rev. Lett.}\ }\textbf {\bibinfo {volume} {101}},\ \bibinfo {pages}
  {147201} (\bibinfo {year} {2008})}\BibitemShut {NoStop}%
\bibitem [{\citenamefont {Rammer}\ and\ \citenamefont
  {Smith}(1986)}]{RammerSmith}%
  \BibitemOpen
  \bibfield  {author} {\bibinfo {author} {\bibfnamefont {J.}~\bibnamefont
  {Rammer}}\ and\ \bibinfo {author} {\bibfnamefont {H.}~\bibnamefont {Smith}},\
  }\href {\doibase 10.1103/RevModPhys.58.323} {\bibfield  {journal} {\bibinfo
  {journal} {Rev. Mod. Phys.}\ }\textbf {\bibinfo {volume} {58}},\ \bibinfo
  {pages} {323} (\bibinfo {year} {1986})}\BibitemShut {NoStop}%
\bibitem [{\citenamefont {Larkin}\ and\ \citenamefont
  {Ovchinnikov}(1986)}]{LO}%
  \BibitemOpen
  \bibfield  {author} {\bibinfo {author} {\bibfnamefont {A.~I.}\ \bibnamefont
  {Larkin}}\ and\ \bibinfo {author} {\bibfnamefont {Y.~N.}\ \bibnamefont
  {Ovchinnikov}},\ }\enquote {\bibinfo {title} {Nonequilibrium
  superconductivity},}\ \ (\bibinfo  {publisher} {Elsevier},\ \bibinfo
  {address} {Amsterdam},\ \bibinfo {year} {1986})\BibitemShut {NoStop}%
\bibitem [{\citenamefont {Belzig}\ \emph {et~al.}(1999)\citenamefont {Belzig},
  \citenamefont {Wilhelm}, \citenamefont {Bruder}, \citenamefont {Schön},\ and\
  \citenamefont {Zaikin}}]{BelzigRev}%
  \BibitemOpen
  \bibfield  {author} {\bibinfo {author} {\bibfnamefont {W.}~\bibnamefont
  {Belzig}}, \bibinfo {author} {\bibfnamefont {F.~K.}\ \bibnamefont {Wilhelm}},
  \bibinfo {author} {\bibfnamefont {C.}~\bibnamefont {Bruder}}, \bibinfo
  {author} {\bibfnamefont {G.}~\bibnamefont {Schön}}, \ and\ \bibinfo {author}
  {\bibfnamefont {A.~D.}\ \bibnamefont {Zaikin}},\ }\href {\doibase
  10.1006/spmi.1999.0710} {\bibfield  {journal} {\bibinfo  {journal}
  {Superlattices and Microstructures}\ }\textbf {\bibinfo {volume} {25}},\
  \bibinfo {pages} {1251 } (\bibinfo {year} {1999})}\BibitemShut {NoStop}%
\bibitem [{\citenamefont {Kopnin}(2001)}]{Kopnin}%
  \BibitemOpen
  \bibfield  {author} {\bibinfo {author} {\bibfnamefont {N.~B.}\ \bibnamefont
  {Kopnin}},\ }\href@noop {} {\emph {\bibinfo {title} {Theory of Nonequilibrium
  Superconductivity}}},\ The International Series of Monographs on Physics\
  (\bibinfo  {publisher} {Clarendon Press},\ \bibinfo {address} {Oxford, UK},\
  \bibinfo {year} {2001})\BibitemShut {NoStop}%
\bibitem [{\citenamefont {Usadel}(1970)}]{Usadel}%
  \BibitemOpen
  \bibfield  {author} {\bibinfo {author} {\bibfnamefont {K.~D.}\ \bibnamefont
  {Usadel}},\ }\href {\doibase 10.1103/PhysRevLett.25.507} {\bibfield
  {journal} {\bibinfo  {journal} {Phys. Rev. Lett.}\ }\textbf {\bibinfo
  {volume} {25}},\ \bibinfo {pages} {507} (\bibinfo {year} {1970})}\BibitemShut
  {NoStop}%
\bibitem [{\citenamefont {Ivanov}\ and\ \citenamefont
  {Fominov}(2006)}]{IvanovFomin}%
  \BibitemOpen
  \bibfield  {author} {\bibinfo {author} {\bibfnamefont {D.~A.}\ \bibnamefont
  {Ivanov}}\ and\ \bibinfo {author} {\bibfnamefont {Y.~V.}\ \bibnamefont
  {Fominov}},\ }\href {\doibase 10.1103/PhysRevB.73.214524} {\bibfield
  {journal} {\bibinfo  {journal} {Phys. Rev. B}\ }\textbf {\bibinfo {volume}
  {73}},\ \bibinfo {pages} {214524} (\bibinfo {year} {2006})}\BibitemShut
  {NoStop}%
\bibitem [{\citenamefont {Machon}\ \emph {et~al.}(2013)\citenamefont {Machon},
  \citenamefont {Eschrig},\ and\ \citenamefont {Belzig}}]{EschrigBC13}%
  \BibitemOpen
  \bibfield  {author} {\bibinfo {author} {\bibfnamefont {P.}~\bibnamefont
  {Machon}}, \bibinfo {author} {\bibfnamefont {M.}~\bibnamefont {Eschrig}}, \
  and\ \bibinfo {author} {\bibfnamefont {W.}~\bibnamefont {Belzig}},\ }\href
  {\doibase 10.1103/PhysRevLett.110.047002} {\bibfield  {journal} {\bibinfo
  {journal} {Phys. Rev. Lett.}\ }\textbf {\bibinfo {volume} {110}},\ \bibinfo
  {pages} {047002} (\bibinfo {year} {2013})}\BibitemShut {NoStop}%
\bibitem [{\citenamefont {Machon}\ \emph {et~al.}(2014)\citenamefont {Machon},
  \citenamefont {Eschrig},\ and\ \citenamefont {Belzig}}]{EschrigBC13a}%
  \BibitemOpen
  \bibfield  {author} {\bibinfo {author} {\bibfnamefont {P.}~\bibnamefont
  {Machon}}, \bibinfo {author} {\bibfnamefont {M.}~\bibnamefont {Eschrig}}, \
  and\ \bibinfo {author} {\bibfnamefont {W.}~\bibnamefont {Belzig}},\ }\href
  {http://stacks.iop.org/1367-2630/16/i=7/a=073002} {\bibfield  {journal}
  {\bibinfo  {journal} {New Journal of Physics}\ }\textbf {\bibinfo {volume}
  {16}},\ \bibinfo {pages} {073002} (\bibinfo {year} {2014})}\BibitemShut
  {NoStop}%
\bibitem [{\citenamefont {{Eschrig}}\ \emph {et~al.}(2015)\citenamefont
  {{Eschrig}}, \citenamefont {{Cottet}}, \citenamefont {{Belzig}},\ and\
  \citenamefont {{Linder}}}]{EschrigBC15}%
  \BibitemOpen
  \bibfield  {author} {\bibinfo {author} {\bibfnamefont {M.}~\bibnamefont
  {{Eschrig}}}, \bibinfo {author} {\bibfnamefont {A.}~\bibnamefont {{Cottet}}},
  \bibinfo {author} {\bibfnamefont {W.}~\bibnamefont {{Belzig}}}, \ and\
  \bibinfo {author} {\bibfnamefont {J.}~\bibnamefont {{Linder}}},\ }\href@noop
  {} {\bibfield  {journal} {\bibinfo  {journal} {ArXiv e-prints}\ } (\bibinfo
  {year} {2015})},\ \Eprint {http://arxiv.org/abs/1504.06258} {arXiv:1504.06258
  [cond-mat.supr-con]} \BibitemShut {NoStop}%
\bibitem [{\citenamefont {Bergeret}\ \emph {et~al.}(2012)\citenamefont
  {Bergeret}, \citenamefont {Verso},\ and\ \citenamefont
  {Volkov}}]{Bergeret12b}%
  \BibitemOpen
  \bibfield  {author} {\bibinfo {author} {\bibfnamefont {F.~S.}\ \bibnamefont
  {Bergeret}}, \bibinfo {author} {\bibfnamefont {A.}~\bibnamefont {Verso}}, \
  and\ \bibinfo {author} {\bibfnamefont {A.~F.}\ \bibnamefont {Volkov}},\
  }\href {\doibase 10.1103/PhysRevB.86.214516} {\bibfield  {journal} {\bibinfo
  {journal} {Phys. Rev. B}\ }\textbf {\bibinfo {volume} {86}},\ \bibinfo
  {pages} {214516} (\bibinfo {year} {2012})}\BibitemShut {NoStop}%
\bibitem [{\citenamefont {Chandra}\ \emph {et~al.}(1990)\citenamefont
  {Chandra}, \citenamefont {Coleman},\ and\ \citenamefont {Larkin}}]{Larkin90}%
  \BibitemOpen
  \bibfield  {author} {\bibinfo {author} {\bibfnamefont {P.}~\bibnamefont
  {Chandra}}, \bibinfo {author} {\bibfnamefont {P.}~\bibnamefont {Coleman}}, \
  and\ \bibinfo {author} {\bibfnamefont {A.~I.}\ \bibnamefont {Larkin}},\
  }\href {http://stacks.iop.org/0953-8984/2/i=39/a=008} {\bibfield  {journal}
  {\bibinfo  {journal} {Journal of Physics: Condensed Matter}\ }\textbf
  {\bibinfo {volume} {2}},\ \bibinfo {pages} {7933} (\bibinfo {year}
  {1990})}\BibitemShut {NoStop}%
\bibitem [{\citenamefont {Nogueira}\ and\ \citenamefont
  {Bennemann}(2004)}]{Flavio04}%
  \BibitemOpen
  \bibfield  {author} {\bibinfo {author} {\bibfnamefont {F.~S.}\ \bibnamefont
  {Nogueira}}\ and\ \bibinfo {author} {\bibfnamefont {K.-H.}\ \bibnamefont
  {Bennemann}},\ }\href {http://stacks.iop.org/0295-5075/67/i=4/a=620}
  {\bibfield  {journal} {\bibinfo  {journal} {EPL (Europhysics Letters)}\
  }\textbf {\bibinfo {volume} {67}},\ \bibinfo {pages} {620} (\bibinfo {year}
  {2004})}\BibitemShut {NoStop}%
\bibitem [{\citenamefont {Chass\'e}\ and\ \citenamefont
  {Tremblay}(2010)}]{Tremblay10}%
  \BibitemOpen
  \bibfield  {author} {\bibinfo {author} {\bibfnamefont {D.}~\bibnamefont
  {Chass\'e}}\ and\ \bibinfo {author} {\bibfnamefont {A.-M.~S.}\ \bibnamefont
  {Tremblay}},\ }\href {\doibase 10.1103/PhysRevB.81.115102} {\bibfield
  {journal} {\bibinfo  {journal} {Phys. Rev. B}\ }\textbf {\bibinfo {volume}
  {81}},\ \bibinfo {pages} {115102} (\bibinfo {year} {2010})}\BibitemShut
  {NoStop}%
\bibitem [{\citenamefont {Moor}\ \emph {et~al.}(2012)\citenamefont {Moor},
  \citenamefont {Volkov},\ and\ \citenamefont {Efetov}}]{Moor12}%
  \BibitemOpen
  \bibfield  {author} {\bibinfo {author} {\bibfnamefont {A.}~\bibnamefont
  {Moor}}, \bibinfo {author} {\bibfnamefont {A.~F.}\ \bibnamefont {Volkov}}, \
  and\ \bibinfo {author} {\bibfnamefont {K.~B.}\ \bibnamefont {Efetov}},\
  }\href {\doibase 10.1103/PhysRevB.85.014523} {\bibfield  {journal} {\bibinfo
  {journal} {Phys. Rev. B}\ }\textbf {\bibinfo {volume} {85}},\ \bibinfo
  {pages} {014523} (\bibinfo {year} {2012})}\BibitemShut {NoStop}%
\bibitem [{\citenamefont {Moor}\ \emph {et~al.}(2015)\citenamefont {Moor},
  \citenamefont {Volkov},\ and\ \citenamefont
  {Efetov}}]{Moor_Volkov_Efetov_SUST}%
  \BibitemOpen
  \bibfield  {author} {\bibinfo {author} {\bibfnamefont {A.}~\bibnamefont
  {Moor}}, \bibinfo {author} {\bibfnamefont {A.~F.}\ \bibnamefont {Volkov}}, \
  and\ \bibinfo {author} {\bibfnamefont {K.~B.}\ \bibnamefont {Efetov}},\
  }\href {http://stacks.iop.org/0953-2048/28/i=2/a=025011} {\bibfield
  {journal} {\bibinfo  {journal} {Superconductor Science and Technology}\
  }\textbf {\bibinfo {volume} {28}},\ \bibinfo {pages} {025011} (\bibinfo
  {year} {2015})}\BibitemShut {NoStop}%
\bibitem [{\citenamefont {Nakata}\ \emph {et~al.}(2014)\citenamefont {Nakata},
  \citenamefont {van Hoogdalem}, \citenamefont {Simon},\ and\ \citenamefont
  {Loss}}]{Nakata_et_al_2014}%
  \BibitemOpen
  \bibfield  {author} {\bibinfo {author} {\bibfnamefont {K.}~\bibnamefont
  {Nakata}}, \bibinfo {author} {\bibfnamefont {K.~A.}\ \bibnamefont {van
  Hoogdalem}}, \bibinfo {author} {\bibfnamefont {P.}~\bibnamefont {Simon}}, \
  and\ \bibinfo {author} {\bibfnamefont {D.}~\bibnamefont {Loss}},\ }\href
  {\doibase 10.1103/PhysRevB.90.144419} {\bibfield  {journal} {\bibinfo
  {journal} {Phys. Rev. B}\ }\textbf {\bibinfo {volume} {90}},\ \bibinfo
  {pages} {144419} (\bibinfo {year} {2014})}\BibitemShut {NoStop}%
\bibitem [{\citenamefont {Chen}\ \emph {et~al.}(2014)\citenamefont {Chen},
  \citenamefont {Horsch},\ and\ \citenamefont
  {Manske}}]{Chen_Horsch_Manske_2014}%
  \BibitemOpen
  \bibfield  {author} {\bibinfo {author} {\bibfnamefont {W.}~\bibnamefont
  {Chen}}, \bibinfo {author} {\bibfnamefont {P.}~\bibnamefont {Horsch}}, \ and\
  \bibinfo {author} {\bibfnamefont {D.}~\bibnamefont {Manske}},\ }\href
  {\doibase 10.1103/PhysRevB.89.064427} {\bibfield  {journal} {\bibinfo
  {journal} {Phys. Rev. B}\ }\textbf {\bibinfo {volume} {89}},\ \bibinfo
  {pages} {064427} (\bibinfo {year} {2014})}\BibitemShut {NoStop}%
\bibitem [{\citenamefont {Buzdin}(2008)}]{Buzdin08}%
  \BibitemOpen
  \bibfield  {author} {\bibinfo {author} {\bibfnamefont {A.}~\bibnamefont
  {Buzdin}},\ }\href {\doibase 10.1103/PhysRevLett.101.107005} {\bibfield
  {journal} {\bibinfo  {journal} {Phys. Rev. Lett.}\ }\textbf {\bibinfo
  {volume} {101}},\ \bibinfo {pages} {107005} (\bibinfo {year}
  {2008})}\BibitemShut {NoStop}%
\bibitem [{\citenamefont {Sickinger}\ \emph {et~al.}(2012)\citenamefont
  {Sickinger}, \citenamefont {Lipman}, \citenamefont {Weides}, \citenamefont
  {Mints}, \citenamefont {Kohlstedt}, \citenamefont {Koelle}, \citenamefont
  {Kleiner},\ and\ \citenamefont {Goldobin}}]{Goldobin12}%
  \BibitemOpen
  \bibfield  {author} {\bibinfo {author} {\bibfnamefont {H.}~\bibnamefont
  {Sickinger}}, \bibinfo {author} {\bibfnamefont {A.}~\bibnamefont {Lipman}},
  \bibinfo {author} {\bibfnamefont {M.}~\bibnamefont {Weides}}, \bibinfo
  {author} {\bibfnamefont {R.~G.}\ \bibnamefont {Mints}}, \bibinfo {author}
  {\bibfnamefont {H.}~\bibnamefont {Kohlstedt}}, \bibinfo {author}
  {\bibfnamefont {D.}~\bibnamefont {Koelle}}, \bibinfo {author} {\bibfnamefont
  {R.}~\bibnamefont {Kleiner}}, \ and\ \bibinfo {author} {\bibfnamefont
  {E.}~\bibnamefont {Goldobin}},\ }\href {\doibase
  10.1103/PhysRevLett.109.107002} {\bibfield  {journal} {\bibinfo  {journal}
  {Phys. Rev. Lett.}\ }\textbf {\bibinfo {volume} {109}},\ \bibinfo {pages}
  {107002} (\bibinfo {year} {2012})}\BibitemShut {NoStop}%
\bibitem [{\citenamefont {Goldobin}\ \emph {et~al.}(2013)\citenamefont
  {Goldobin}, \citenamefont {Kleiner}, \citenamefont {Koelle},\ and\
  \citenamefont {Mints}}]{Goldobin13}%
  \BibitemOpen
  \bibfield  {author} {\bibinfo {author} {\bibfnamefont {E.}~\bibnamefont
  {Goldobin}}, \bibinfo {author} {\bibfnamefont {R.}~\bibnamefont {Kleiner}},
  \bibinfo {author} {\bibfnamefont {D.}~\bibnamefont {Koelle}}, \ and\ \bibinfo
  {author} {\bibfnamefont {R.~G.}\ \bibnamefont {Mints}},\ }\href {\doibase
  10.1103/PhysRevLett.111.057004} {\bibfield  {journal} {\bibinfo  {journal}
  {Phys. Rev. Lett.}\ }\textbf {\bibinfo {volume} {111}},\ \bibinfo {pages}
  {057004} (\bibinfo {year} {2013})}\BibitemShut {NoStop}%
\bibitem [{\citenamefont {Moor}\ \emph {et~al.}(2013)\citenamefont {Moor},
  \citenamefont {Volkov},\ and\ \citenamefont {Efetov}}]{Moor13}%
  \BibitemOpen
  \bibfield  {author} {\bibinfo {author} {\bibfnamefont {A.}~\bibnamefont
  {Moor}}, \bibinfo {author} {\bibfnamefont {A.~F.}\ \bibnamefont {Volkov}}, \
  and\ \bibinfo {author} {\bibfnamefont {K.~B.}\ \bibnamefont {Efetov}},\
  }\href {\doibase 10.1103/PhysRevB.87.100504} {\bibfield  {journal} {\bibinfo
  {journal} {Phys. Rev. B}\ }\textbf {\bibinfo {volume} {87}},\ \bibinfo
  {pages} {100504(R)} (\bibinfo {year} {2013})}\BibitemShut {NoStop}%
\bibitem [{\citenamefont {Bauer}\ \emph {et~al.}(2004)\citenamefont {Bauer},
  \citenamefont {Bentner}, \citenamefont {Aprili}, \citenamefont {Della~Rocca},
  \citenamefont {Reinwald}, \citenamefont {Wegscheider},\ and\ \citenamefont
  {Strunk}}]{Bauer_et_al_2004}%
  \BibitemOpen
  \bibfield  {author} {\bibinfo {author} {\bibfnamefont {A.}~\bibnamefont
  {Bauer}}, \bibinfo {author} {\bibfnamefont {J.}~\bibnamefont {Bentner}},
  \bibinfo {author} {\bibfnamefont {M.}~\bibnamefont {Aprili}}, \bibinfo
  {author} {\bibfnamefont {M.~L.}\ \bibnamefont {Della~Rocca}}, \bibinfo
  {author} {\bibfnamefont {M.}~\bibnamefont {Reinwald}}, \bibinfo {author}
  {\bibfnamefont {W.}~\bibnamefont {Wegscheider}}, \ and\ \bibinfo {author}
  {\bibfnamefont {C.}~\bibnamefont {Strunk}},\ }\href {\doibase
  10.1103/PhysRevLett.92.217001} {\bibfield  {journal} {\bibinfo  {journal}
  {Phys. Rev. Lett.}\ }\textbf {\bibinfo {volume} {92}},\ \bibinfo {pages}
  {217001} (\bibinfo {year} {2004})}\BibitemShut {NoStop}%
\bibitem [{\citenamefont {Grein}\ \emph {et~al.}(2009)\citenamefont {Grein},
  \citenamefont {Eschrig}, \citenamefont {Metalidis},\ and\ \citenamefont
  {Sch\"on}}]{Grein_et_al_2009}%
  \BibitemOpen
  \bibfield  {author} {\bibinfo {author} {\bibfnamefont {R.}~\bibnamefont
  {Grein}}, \bibinfo {author} {\bibfnamefont {M.}~\bibnamefont {Eschrig}},
  \bibinfo {author} {\bibfnamefont {G.}~\bibnamefont {Metalidis}}, \ and\
  \bibinfo {author} {\bibfnamefont {G.}~\bibnamefont {Sch\"on}},\ }\href
  {\doibase 10.1103/PhysRevLett.102.227005} {\bibfield  {journal} {\bibinfo
  {journal} {Phys. Rev. Lett.}\ }\textbf {\bibinfo {volume} {102}},\ \bibinfo
  {pages} {227005} (\bibinfo {year} {2009})}\BibitemShut {NoStop}%
\bibitem [{\citenamefont {Braude}\ and\ \citenamefont
  {Nazarov}(2007)}]{Braude07}%
  \BibitemOpen
  \bibfield  {author} {\bibinfo {author} {\bibfnamefont {V.}~\bibnamefont
  {Braude}}\ and\ \bibinfo {author} {\bibfnamefont {Y.~V.}\ \bibnamefont
  {Nazarov}},\ }\href {\doibase 10.1103/PhysRevLett.98.077003} {\bibfield
  {journal} {\bibinfo  {journal} {Phys. Rev. Lett.}\ }\textbf {\bibinfo
  {volume} {98}},\ \bibinfo {pages} {077003} (\bibinfo {year}
  {2007})}\BibitemShut {NoStop}%
\bibitem [{\citenamefont {{Muduli}}\ \emph {et~al.}(2014)\citenamefont
  {{Muduli}}, \citenamefont {{Wang}}, \citenamefont {{Zhao}},\ and\
  \citenamefont {{Blamire}}}]{Muduli_et_al_arXiv_2015}%
  \BibitemOpen
  \bibfield  {author} {\bibinfo {author} {\bibfnamefont {P.~K.}\ \bibnamefont
  {{Muduli}}}, \bibinfo {author} {\bibfnamefont {X.~L.}\ \bibnamefont
  {{Wang}}}, \bibinfo {author} {\bibfnamefont {J.~H.}\ \bibnamefont {{Zhao}}},
  \ and\ \bibinfo {author} {\bibfnamefont {M.~G.}\ \bibnamefont {{Blamire}}},\
  }\href@noop {} {\bibfield  {journal} {\bibinfo  {journal} {ArXiv e-prints}\ }
  (\bibinfo {year} {2014})},\ \Eprint {http://arxiv.org/abs/1410.6741}
  {arXiv:1410.6741 [cond-mat.mes-hall]} \BibitemShut {NoStop}%
\bibitem [{\citenamefont {{Eschrig}}(2015)}]{Eschrig_Review_arXiv_2015}%
  \BibitemOpen
  \bibfield  {author} {\bibinfo {author} {\bibfnamefont {M.}~\bibnamefont
  {{Eschrig}}},\ }\href@noop {} {\bibfield  {journal} {\bibinfo  {journal}
  {ArXiv e-prints}\ } (\bibinfo {year} {2015})},\ \Eprint
  {http://arxiv.org/abs/1509.02242} {arXiv:1509.02242 [cond-mat.supr-con]}
  \BibitemShut {NoStop}%
\bibitem [{\citenamefont {Likharev}(1979)}]{Likharev}%
  \BibitemOpen
  \bibfield  {author} {\bibinfo {author} {\bibfnamefont {K.~K.}\ \bibnamefont
  {Likharev}},\ }\href {\doibase 10.1103/RevModPhys.51.101} {\bibfield
  {journal} {\bibinfo  {journal} {Rev. Mod. Phys.}\ }\textbf {\bibinfo {volume}
  {51}},\ \bibinfo {pages} {101} (\bibinfo {year} {1979})}\BibitemShut
  {NoStop}%
\end{thebibliography}

%

\end{document}